\documentclass[iop]{emulateapj}

\usepackage{natbib}
\usepackage{longtable}
\shorttitle{Revisiting Stochastic Variability of AGNs with Structure Functions}
\shortauthors{Szymon~Koz{\l}owski}

\begin{document}

\title{Revisiting Stochastic Variability of AGNs with Structure Functions}

\author{Szymon~Koz{\l}owski\altaffilmark{1}}

\altaffiltext{1}{Warsaw University Observatory, Al. Ujazdowskie 4, 00-478 Warszawa, Poland; simkoz@astrouw.edu.pl}


\begin{abstract}
Discrepancies between reported structure function (SF) slopes and their overall flatness as compared to expectations from the damped random walk (DRW) model,
which generally well describes the variability of active galactic nuclei (AGNs), have triggered us to study this problem in detail. 
We review common AGN variability observables and identify their most common problems.
Equipped with this knowledge, we study $\sim$9000 $r$-band AGN light curves from Stripe 82 of the Sloan Digital Sky Survey,
using SFs described by stochastic processes with the power exponential covariance matrix of the signal. 
We model the ``subensemble'' SFs in the redshift--absolute magnitude bins with the full SF equation (including the turnover and the noise part) 
and a single power law (SPL; in the ``red noise regime'' after subtracting the noise term). 
The distribution of full-equation SF (SPL) slopes
peaks at $\gamma=0.55\pm 0.08$ ($0.52\pm 0.06$) and is consistent with the DRW model. 
There is a hint of a weak correlation of $\gamma$ with the luminosity and a lack of correlation with the black hole mass.
The typical decorrelation timescale in the optical is $\tau=0.97\pm0.46$ year.
The SF amplitude at one year obtained from the SPL fitting is $SF_0=0.22\pm0.06$ mag and is overestimated because the SF is already at the turnover part,
so the true value is $SF_0=0.20\pm0.06$ mag. The asymptotic variability is $SF_\infty=0.25\pm0.06$ mag. It
is strongly anticorrelated with both the luminosity and the Eddington ratio and is correlated with the black hole mass.
The reliability of these results is fortified with Monte Carlo simulations.
\end{abstract}

\keywords{accretion, accretion disks -- galaxies: active -- methods: data analysis -- quasars: general}


\section{Introduction}

One of the flagship properties of active galactic nuclei (AGNs) is their variability. 
It was discovered (\citealt{1963ApJ...138...30M,1963Natur.198..650S}) right after or along with the discovery of this class of objects in 1963 (\citealt{1963Natur.197.1040S,1963Natur.197.1041G}).
Because AGNs are the most powerful sources of continuous light in the universe (with the bolometric luminosities as high as $10^{48}$~erg~s$^{-1}$ or $10^{14} L_\odot$), their variability
at timescales of months to years of order of 10\% of the total light must be enormous. 
There are multiple lines of evidence that AGNs are powered by the accretion of matter onto supermassive black holes (SMBHs; \citealt{1973A&A....24..337S,1984ARA&A..22..471R}).
For example, chromatic microlensing of strongly gravitationally lensed AGNs provides us with
the sizes ($\sim1$ au in optical) and temperature profiles of accretion disks (\citealt{1994MNRAS.268..135C,2004ApJ...605...58K,2010ApJ...712.1129M,2010ApJ...709..278D,2011ApJ...729...34B}).

During the five decades since their discovery, 
we have witnessed an increasing number of studies on the variability of AGNs,
on both data analyses and theory. The theoretical studies examined accretion disk instabilities, surface temperature fluctuations, and variable heating from coronal X-rays (e.g., \citealt{1976MNRAS.175..613S,1993A&A...272....8R,1995ApJ...441..354C,1998ApJ...504..671K,2014ApJ...783..105R}) but also
currently considered as non-viable: microlensing and chain supernovae, also known as the starburst model (\citealt{1993Natur.366..242H,1995ApJ...444L..13B,1998ApJ...504..671K}).
On the data side, aperiodic luminosity fluctuations have been described by means of the rms variability (sometimes the squared rms) as a function
of the time difference, or equivalently frequency, between epochs, and these methods are known as the structure function (SF) and
power spectral density (PSD) analyses, respectively. Continuously growing, in length and cadence, quasar light curves from optical surveys
enabled even more detailed studies and quantification of AGN variability that include direct light curve modeling. A model that works particularly 
well is the damped random walk model
(DRW; \citealt{2009ApJ...698..895K,2010ApJ...708..927K,2010ApJ...721.1014M,2011AJ....141...93B,2011ApJ...728...26M,2011ApJ...735...80Z,2012ApJ...753..106M,2012ApJ...760...51R,2013ApJ...765..106Z}), 
although various alternatives have been tested (e.g., \citealt{2011ApJ...730...52K,2013ApJ...765..106Z,2014MNRAS.445.3055P}), and some
apparent deviations from it have been reported (\citealt{2011ApJ...743L..12M,2015MNRAS.451.4328K}).
DRW is also the simplest of a broader, more general class of continuous-time autoregressive moving average (CARMA) models, recently considered by \cite{2014ApJ...788...33K} and
\cite{2016A&A...585A.129S}.

\begin{figure*}
\centering
\includegraphics[width=16.0cm]{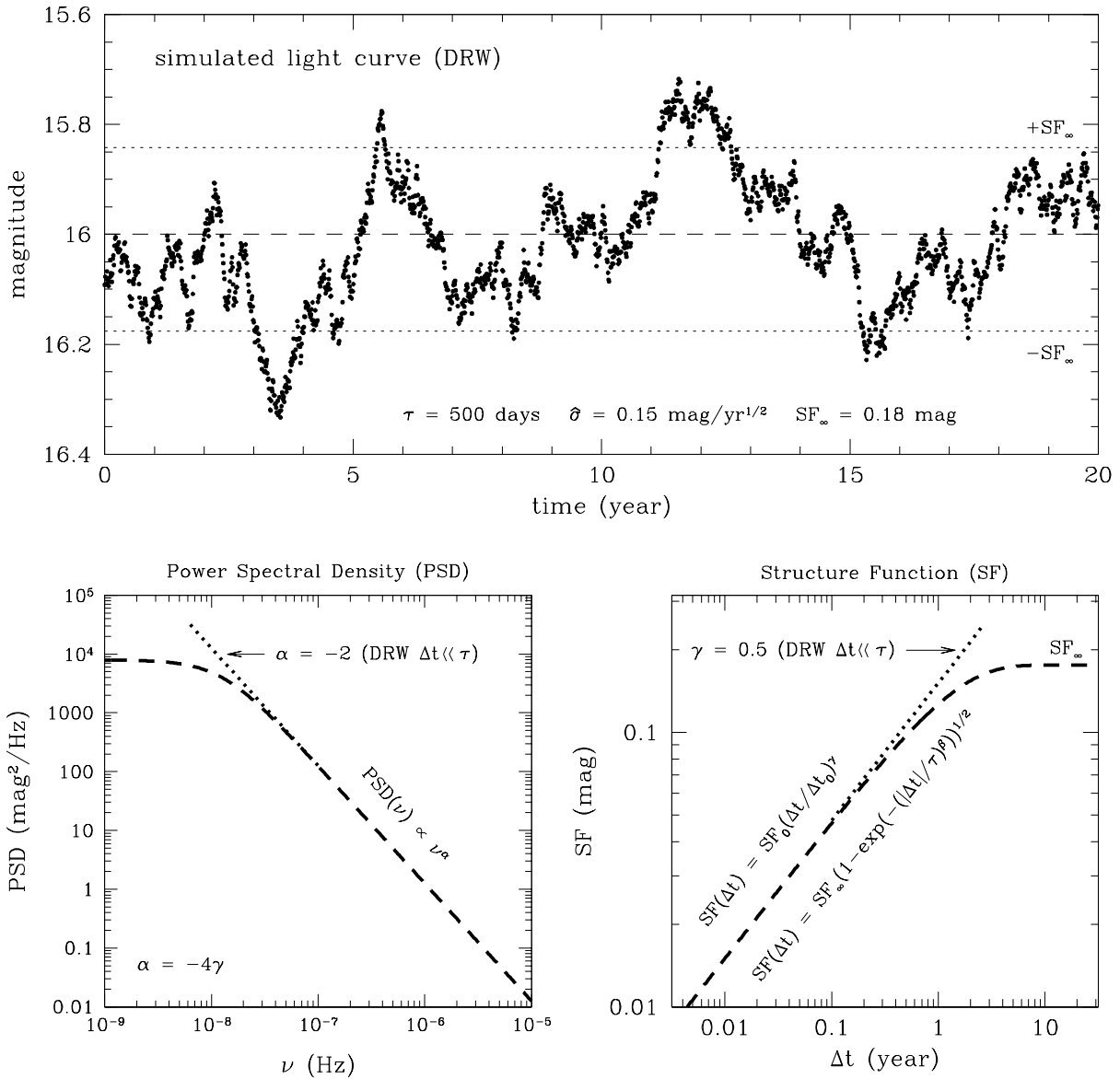}
\caption{Presentation of basic AGN variability concepts and measures.
Top panel: an idealized simulated AGN light curve using DRW is shown. It can be either directly fit by the DRW (or CARMA) model and return the model parameters or
studied via the PSD (bottom left panel) or SF (bottom right panel) analysis. Each panel also presents basic variability features
related to these AGN variability measures.}
\label{fig:vardef}
\end{figure*}


The first goal of this paper is to {\it explicitly} review AGN variability observables, in particular 
their connection to the core of the underlying process. 
We will be primarily interested in the SF analysis, but we also skim the PSD analysis, both 
typically considered for sparse or short light curves. We briefly review the DRW method (or its siblings, e.g., \citealt{2013ApJ...765..106Z}) used to model light curves
(Figure~{\ref{fig:vardef}). The importance of the first two of these observables lies in their model-independent
capability to directly measure the covariance function of the signal (i.e., pinpoint the underlying process),
while the DRW light curve modeling lacks this capability (\citealt{2016MNRAS.459.2787K}).
The second goal is to apply the SF method to $\sim$9000 quasar light curves from the Sloan Digital Sky Survey (SDSS) Stripe 82 (S82) 
in order to find the covariance function of the process and also basic variability parameters.
Subsequently we will be interested in correlations between the shape of the covariance function and the physical parameters of AGNs, 
such as the black hole mass, luminosity, Eddington ratio, and redshift (or equivalently the rest-frame wavelength).
By means of Monte Carlo simulations, we will show how the SF or DRW is affected by data quality or sampling; in particular
we present the biases in the derived SF and DRW parameters.

The paper is organized as follows. In Section~\ref{sec:var}, we describe and discuss
variability observables for AGNs. In Section~\ref{sec:data}, we analyze the SDSS S82 data with the SF and DRW models 
as well as simulate similar data sets in order to understand any problems, biases, or systematics. 
In Section~\ref{sec:discussion} we discuss our findings.
The paper is summarized in Section~\ref{sec:summary}.


\section{Description of Variability}
\label{sec:var}

We will now explicitly review vital details of measuring AGN variability.
We briefly recollect that AGN variability resembles and is well described by stochastic processes: it is aperiodic,
meaning it has smooth PSDs (with no peaks on periodograms; e.g, \citealt{2009ApJ...698..895K}).
Throughout this paper we will explicitly assume that AGN variability is due to a single stochastic process or a combination of stochastic processes (\citealt{2011ApJ...730...52K}).

\subsection{SF in a Nutshell}

Let us consider a data set composed of a collection of measured data $y_i$ (e.g., magnitudes) at times $t_i$ with $i=1, ..., N$ points 
(hereafter a light curve) that can be represented as a sum of the true signal $s_i$ and noise $n_i$, $y_i=s_i+n_i$
(see \citealt{1981ApJS...45....1S, 1982ApJ...263..835S, 1989ApJ...343..874S,1992ApJ...398..169R,1992ApJ...385..404P,1992ApJ...385..416P} for an intuitive description of a
time series analysis). 
Having an AGN light curve, we will be interested in the relationship of this light curve to its copy shifted by time $\Delta t$ for various $\Delta t=(0, \dots, \infty)$ days, 
because, as we will show, such a relationship contains the key to understanding the origin of variability.
For each time shift $\Delta t$, often called the time ``lag,'' we will find all pairs of points occurring at the same instant with $\Delta t = |t_i-t_j|$, where $i$ and $j$ are
the indices in the original and shifted light curves.
Because AGNs are typically distant sources, they participate in the Hubble flow, and their redshifts $z$ are often substantial. We correct the observed-frame variability to the rest frame via
$\Delta t = |t_i-t_j|(1+z)^{-1}$. Throughout this paper we will assume $\Delta t$ to be the rest-frame time lag.

If an underlying stochastic process leading to variability is stationary (its mean, variance, and probability do not change with time), then 
the relationship between the two light curves (the original and the shifted one) can be quantified by the covariance function 

\begin{eqnarray}
\label{eq:covgen}
{\rm cov}(\Delta t) &\equiv& {\rm cov}(y(t),y(t+\Delta t)) \equiv \nonumber \\
&\equiv& {\rm cov}(y_i, y_j) \equiv \nonumber \\
&\equiv&  \frac{1}{N_{\rm \Delta t~pairs}}\sum_{i=1}^{N_{\rm \Delta t~pairs}}(y_i-\langle y \rangle)(y_j-\langle y\rangle) \equiv \nonumber \\ 
                    &\equiv& \langle(y_i-\langle y\rangle)(y_j-\langle y\rangle)\rangle
\end{eqnarray}
\noindent
and the covariance for $\Delta t=0$ days ($y_i=y_j$) is

\begin{equation}
{\rm cov}(0) \equiv {\rm cov}(y_i, y_i) \equiv {\rm var}(y_i) \equiv \langle(y_i-\langle y\rangle)^2\rangle,
\end{equation}
\noindent
where var$(y_i)$ is the data variance and $\langle y\rangle$ is the mean.

It is straightforward to show (\citealt{1992ApJ...385..404P}) that the covariance of data (from Equation~(\ref{eq:covgen})) can be expressed in terms of the data variance and the SF as

\begin{equation}
\label{eq:sftheory}
{\rm cov}(y_i, y_j) \equiv {\rm var}(y_i) - V(y_i, y_j),
\end{equation}

\noindent 
where  

\begin{equation}
V(y_i, y_j) = \frac{1}{2}\langle(y_i-y_j)^2\rangle
\label{eq:SFtheoryV}
\end{equation}
\noindent 
is the ``theoretical'' SF (in units of squared magnitude), as opposed to the typically reported one that scales as $SF=\sqrt{2V}$ (in units of magnitude).

Throughout this paper, we will be interested in the form of the covariance function of the signal cov$(s_i, s_j)$ 
(or equivalently in the autocorrelation function (ACF), which by definition is $ACF(\Delta t)\equiv{\rm cov}(s_i, s_j)/\sigma_s^2$) 
describing the underlying process leading to variability.
It can be directly obtained from the data through the structure function (from Equation~(\ref{eq:sftheory})):

\begin{eqnarray}
V(y_i, y_j) &=& {\rm var}(y_i) - {\rm cov}(y_i, y_j) = {\rm var}(s_i) + \nonumber\\
            &+& {\rm var}(n_i) - {\rm cov}(s_i, s_j) - {\rm cov}(n_i, n_j) = \nonumber \\
	    &=& \sigma_s^2 + \sigma_n^2 - {\rm cov}(s_i, s_j),
\label{eq:SFtheory}
\end{eqnarray}
\noindent 
where ${\rm var}(s_i)\equiv\sigma_s^2$, ${\rm var}(n_i)\equiv\sigma_n^2$, and ${\rm cov}(s_i, n_i)={\rm cov}(n_i, n_j)\equiv 0$ because 
both the data and noise are assumed to be uncorrelated with the noise. We implicitly assumed above that both the signal and noise have Gaussian distributions (${\rm var}\equiv\sigma^2$);
however, while we will be interested in Gaussian processes for the signal, the noise may have a non-Gaussian distribution (${\rm var}(n_i)\neq\sigma_n^2$).
From Equation~(\ref{eq:SFtheory}) it follows that the rest-frame SF is
\begin{equation}
\label{eq:sfobs}
SF_{\rm obs}(\Delta t) = \sqrt{2\sigma_s^2 + 2\sigma_n^2 - 2{\rm cov}(s_i, s_j)}
\end{equation}
\noindent 
and contains information on the covariance function of the process causing the variability. 
The true underlying SF, after subtracting the noise term $2\sigma_n^2$ or $2{\rm var}(n)$ for a non-Gaussian distribution, is

\begin{equation}
SF_{\rm true}(\Delta t) = \sqrt{SF_{\rm obs}(\Delta t)^2 - 2\sigma_n^2},
\label{eq:sftrue}
\end{equation}
\noindent 
and hence

\begin{eqnarray}
SF_{\rm true}(\Delta t) &=& \sqrt{2\sigma_s^2(1-ACF(\Delta t))} = \nonumber \\
			&=& SF_\infty\sqrt{1-ACF(\Delta t)},
\label{eq:simone}
\end{eqnarray}
\noindent
and \cite{2010ApJ...721.1014M} defines $SF_\infty=\sqrt{2}\sigma_s$ because $ACF(\Delta t)\rightarrow 0$ as $\Delta t \rightarrow \infty$. 
The ACF can take any value between $0$ and $1$, and its shape is simply a
function of $\Delta t$ (Figure~\ref{fig:PE}). It can be also thought of as a ``memory function,'' where $ACF=1$ ($ACF=0$) reflects the perfect (lack of) memory, 
and any values in between 0 and 1 show how strongly two data points separated by $\Delta t$ ``remember'' each other. 

\begin{figure*}
\centering
\includegraphics[width=16.0cm]{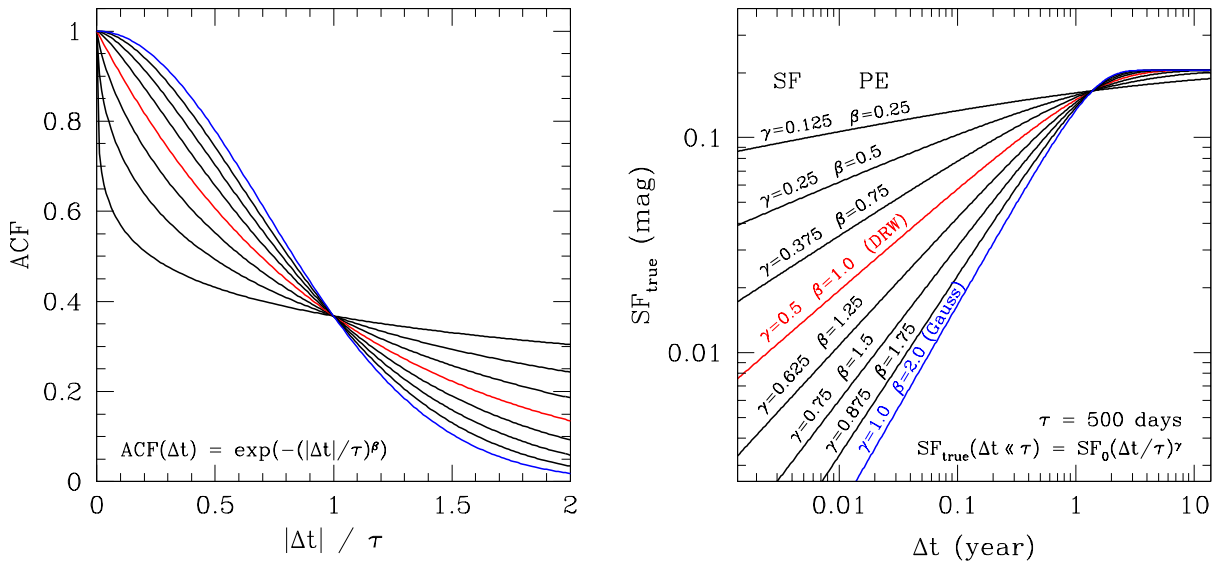}
\caption{Graphic explanation of the connection between the
power exponential ACFs (left panel) and their corresponding SFs (right panel). DRW (marked in red) is the case with $\beta=1$ 
and produces the SF slope $\gamma=0.5$ for $\Delta t \ll \tau$ (i.e., the red noise regime). 
Steeper or shallower SFs than DRW may be due to a combination of DRW processes and can be viewed as having a stronger ($\beta>1$) or weaker ($\beta<1$) ACF, respectively.}
\label{fig:PE}
\end{figure*}

The calculation of $SF_{\rm obs}$ is typically obtained from (Equation~(\ref{eq:SFtheoryV})):
\begin{eqnarray}
SF_{\rm obs}(\Delta t) &=& \sqrt{\frac{1}{N_{\rm \Delta t~pairs}}\sum_{i=1}^{N_{\rm \Delta t~pairs}} (y(t)-y(t+\Delta t))^2} \equiv \nonumber \\
	     &\equiv& {\rm rms}\left[y(t)-y(t+\Delta t)\right]
\label{eq:sfrms}
\end{eqnarray}
\noindent
or equivalently
\begin{eqnarray}
SF_{\rm obs}(\Delta t) &=& 0.741 \times {\rm IQR},
\label{eq:sfrmsMcL}
\end{eqnarray}
\noindent
where IQR is the interquartile range between 25\% and 75\% of the sorted $(y(t)-y(t+\Delta t))$ distribution, and
the 0.741 coefficient is the conversion to $\sigma$ for the normal or Gaussian distribution (\citealt{2012ApJ...753..106M}). 
The latter description provides the rms values less affected by outliers in the distribution or if the distribution is non-Gaussian. 
From Equation~(\ref{eq:sfrms}) it is obvious that the SF measures the amount of the rms variability as a function of the time interval or ``lag'' ($\Delta t$) between points. 

\begin{figure}
\centering
\includegraphics[width=8.2cm]{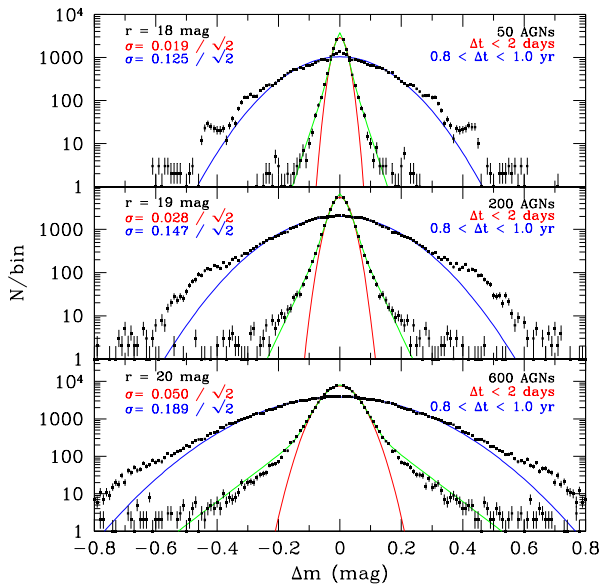}
\caption{Distributions of magnitude differences $\Delta m$  for two $\Delta t$ regimes, $\Delta t<2$ days (points with red or green fits) and
$0.8<\Delta t<1$ year (points with blue fits), and for the AGN mean magnitudes falling into three levels, $r=18\pm0.1$ mag, $19\pm0.05$ mag, and $20\pm 0.05$ mag (panels from top to bottom).
For $\Delta t<2$ days, $SF_{\rm true}\approx 0$ mag, so these distributions, in fact, show true uncertainty distributions (they need to be divided by $\sqrt{2}$) for a given magnitude.
The photometric error bars are weakly non-Gaussian (red Gaussian fits) and are better described as a sum of a Gaussian and an exponential function (green fits).
The distribution of differences at the longer lag is also weakly non-Gaussian (blue Gaussian fits).}
\label{fig:errorbars}
\end{figure}

Note also that the rms around the mean is ${\rm rms^2}[x-\langle x\rangle]=\langle x\rangle^2+\sigma_x^2$,
but in our case $\langle x\rangle=0$. We also calculate the rms of the differences (and not around the mean), so ${\rm rms^2}[x_i-x_j]=2\sigma_x^2$.
Equation~(\ref{eq:sftrue}) becomes then

\begin{eqnarray}
\label{eq:sfrms2}
SF_{\rm true}^2(\Delta t) &\equiv& {\rm rms}^2\left[y(t)-y(t+\Delta t)\right] - \nonumber \\ 
	               &-& {\rm rms}^2\left[y(t)-y(t+(\Delta t\rightarrow 0))\right],
\end{eqnarray}
\noindent
and a variant of it was used to study the SFs of {\it Spitzer} mid-infrared AGNs in \cite{2010ApJ...716..530K,2016ApJ...817..119K} 
(the second term was the rms for the nonvariable field objects without the requirement $\Delta t\rightarrow 0$).

In fact, \cite{1992ApJ...385..404P} seems to provide the most correct way to estimate SFs (although sensitive to outliers) from

\begin{eqnarray}
\label{eq:sftruSch}
SF_{\rm true}^2(\Delta t) &=& \frac{1}{N_{\rm \Delta t~pairs}}\sum_{i=1}^{N_{\rm \Delta t~pairs}} (y(t)-y(t+\Delta t))^2 - \nonumber \\
		         &-& \sigma_n^2(y(t)) - \sigma_n^2(y(t+\Delta t)),
\end{eqnarray}
\noindent 
where each magnitude measurement $y(t)$ is accompanied by its own $\sigma_n(y)$ (i.e., dispersion of noise for this particular magnitude, 
typically estimated via dispersions from light curves of nonvariable field objects of the same magnitude) and must be obtained prior to the SF calculation.


\subsection{The Photometric Noise}

From Equation~(\ref{eq:sfobs}) it follows that the shape of the observed SF as a function of $\Delta t$ strongly depends on $\sigma_n$ 
for $\Delta t\rightarrow 0$ and is a function of a survey's photometric properties and a source magnitude.
The noise term $2\sigma_n^2$ in Equation~(\ref{eq:sftrue}) could be directly calculated from the photometric error bars
by assuming that $G(\sigma_\epsilon)=G(\sigma_n)$, which typically is not true as the measurements' uncertainties 
are often incorrectly estimated (see a discussion by, e.g., \citealt{2016AcA....66....1S}). There exist, operationally, better ways to do it: (1) it can be calculated as the rms
of measurements (or $0.741 \times {\rm IQR}$) for nonvariable field objects with the same magnitude (e.g., \citealt{2010ApJ...716..530K,2016ApJ...817..119K}), 
or (2) it can be directly estimated from  $SF_{\rm obs}(\Delta t)$ for $\Delta t \rightarrow 0$, because only then $SF_{\rm true}(\Delta t)\rightarrow 0$.
Please note that, by setting var$(n)=\sigma_n^2$, we implicitly assume the noise distribution to be Gaussian, but from Figure~\ref{fig:errorbars}
we see that the SDSS S82 AGNs have a weakly non-Gaussian distribution of uncertainties. If the distribution of uncertainties 
can be approximated by any function for which the variance exists, instead of subtracting the noise term $2\sigma_n^2$, one
should subtract $2$var$(n)$.

One may ask if the exponential wings of the distributions shown in Figure~\ref{fig:errorbars} are due to the photometric noise or rather are due to the underlying AGN variability.
As already explained, $SF_{\rm obs}(\Delta t)\approx2$var$(n)$ for $\Delta t \rightarrow 0$, because only then $SF_{\rm true}(\Delta t)\rightarrow 0$.
We have checked the corresponding distributions for the field sources (stars, galaxies) as in Figure~\ref{fig:errorbars} and they show similar exponential wings,
so they are not due to AGN variability.


\subsection{Common SF Issues}


\begin{figure}
\centering
\includegraphics[width=8.2cm]{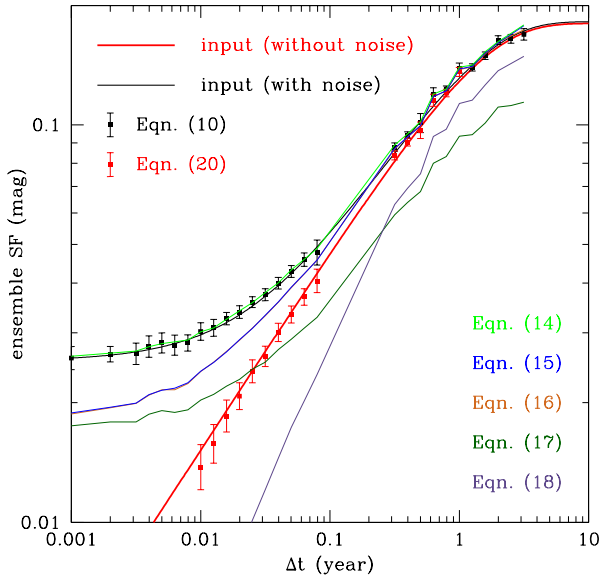}
\caption{Example SF calculations for 1000 simulated AGN light curves in a bin $z=1.6\pm0.1$ with $M_i=-26.0\pm0.25$ mag. In simulations we used the DRW model. 
The input signal (signal and noise) SF is shown as the thick red (black) line. The calculated SFs from the two methods used in this study are given as red and black points. We also show
what common SF measures return (note that they should return the red line).}
\label{fig:SFexplain2}
\end{figure}

One of the triggers for this study were both seemingly flat SFs ($\gamma=0.1$--0.4; as compared to that required by DRW, $\gamma=0.5$) and the discrepancies between them.

It is obvious that SFs (or PSDs, \citealt{2011ApJ...730...52K}) cannot be single power laws (SPL) for $\Delta t \rightarrow\infty$, 
because a very large or infinite power would be required at long timescales.
Therefore a typical SF is rather bent (Figure~\ref{fig:PE}), and it has two (or sometimes three) regimes: for $\Delta t \ll \tau$ (the red noise regime) it can be fit as an SPL of the form
\begin{equation}
SF(\Delta t) = SF_0\left(\frac{\Delta t}{\Delta t_0}\right)^\gamma,
\label{eq:sfsinglefit}
\end{equation}
\noindent
where SF$_0$ is the variability amplitude at a fixed timescale $\Delta t_0$. Around $\Delta t \approx \tau$ we observe a transition, a turnover, from an SPL
into another SPL with $\gamma=0$ (i.e., the white noise), which is the second regime for $\Delta t\rightarrow \infty$. The third regime is for 
$\Delta t\approx 0$ days; in particular if the noise term is not subtracted (or subtracted incompletely), then we are in an another SPL regime with $\gamma=0$ 
(i.e., white noise regime due to the photometric noise).
This very short timescale regime vanishes provided the photometric noise is removed correctly. Because the real data rarely show a Gaussian-like behavior in the noise,
SFs are uncertain in this regime, and it is best to simply exclude them from the analyses.

In early studies, the turnover timescale $\tau$ was not known (which is of about 500 days in the optical, \citealt{2010ApJ...721.1014M,2012ApJ...753..106M}, or about one year from the SFs in this study), 
and SFs were fit as an SPL, which is obviously incorrect in the light of the infinite power required to generate variations at long lags, 
and henceforth the measured SFs must have been flat by definition  (\citealt{2005AJ....129..615D}).

A number of authors use SF definitions that are obviously some measure of AGN variability, but they are invalid in the light of measuring the underlying true SF and hence ACF.
The common problem is what kind of noise one should subtract when fitting an SPL SF.
The rule of thumb should be as follows: for a $(m_i-m_j)^2$ term one subtracts $2\sigma_n^2$ from it, 
and for a $(m_i-\langle m\rangle)^2$ term one subtracts just a single $\sigma_n^2$.
\cite{2014MNRAS.439..703G}, apart from introducing a new method of variability analysis with the Slepian wavelets variance, combines a number of SF calculation procedures found in the literature. 
We have tested these SF definitions on the simulated data (see Figure~\ref{fig:SFexplain2}):

\begin{equation}
SF(\Delta t) = \langle(m_i-m_j)^2\rangle
\end{equation}
\noindent
from \cite{1984ApJ...284..126S} and \cite{1994MNRAS.268..305H} is in fact twice the definition of the ``theoretical'' SF from Equation~(\ref{eq:SFtheoryV}). It does 
not subtract the noise term, and the square root of this equation (as in \citealt{2005AJ....129..615D} and \citealt{2014ApJ...782...37C}) is equivalent to Equation~(\ref{eq:sfrmsMcL}). Fitting 
an SPL SF to it gives flat SF slopes and cannot be used to infer the ACF. One would need to fit a full, four-parameter SF to obtain the correct result.

\begin{equation}
SF(\Delta t) = \sqrt{\langle(m_i-m_j)^2\rangle - \langle\sigma^2\rangle }
\end{equation}
\noindent
from \cite{2009ApJ...696.1241B} subtracts an incomplete noise term (it should have a 2 in front of the noise term) and leads to flatter SFs (see Figure~\ref{fig:SFexplain2}).

\begin{equation}
SF(\Delta t) = \sqrt{\frac{\pi}{2}\langle|m_i-m_j|\rangle^2 - \langle\sigma^2\rangle }
\end{equation}
\noindent
from \cite{1996ApJ...463..466D}, \cite{2004ApJ...601..692V}, and \cite{2009ApJ...696.1241B} subtracts an incomplete noise term (it should have a 2 in front of the noise term) and leads to flatter SFs.

\begin{equation}
SF(\Delta t) = {\rm median}[(m_i-m_j)^2]
\end{equation}
\noindent
from \cite{2005MNRAS.356..331S} does not subtract the noise term and returns flat SFs. In fact, the square root of it must be taken, and then it returns an SF shape 
to that from Equation~(\ref{eq:sfrmsMcL}), but the whole SF is shifted toward lower values.

\begin{equation}
SF(\Delta t) = \left<\sqrt{\frac{\pi}{2}}|m_i-m_j| - \sqrt{\sigma_i^2+\sigma_j^2}\right>_{\Delta t}
\end{equation}
\noindent
from \cite{2010ApJ...714.1194S} is a strange measure; it generally looks as if it was subtracting too much of the noise term, leading to steeper and overall lower SF values.
A plethora of SF definitions cause problems in their interpretations or comparisons; for example, both \cite{2010ApJ...721.1941S} and \cite{2014ApJ...782..119M} issue errata on SF definitions.
\cite{2010MNRAS.404..931E} provide an overview of an SF analysis and in particular point out problems (unexpected breaks or wiggles) and caveats in its interpretation.


\subsection{SF Calculations and Fitting}
\label{sec:sfcalcfit}

In this paper, we will be using two methods to estimate and model SFs:

(1) In the first one, we will use Equation~(\ref{eq:sfrmsMcL}), that is, without the photometric noise subtraction, to calculate the observed SF. We will fit to it the full, four-parameter SF
described by the power exponential (PE) covariance matrix of the form $\sigma_s^2 \exp(-(|\Delta t|/\tau)^\beta)$.
The four parameters include the power $\beta$, decorrelation timescale $\tau$, 
variance at long timescales $SF_\infty$, and the noise term $\sigma_n$:

\begin{equation}
SF_{\rm obs}^2(\Delta t)=SF_\infty^2\left(1-\exp(-(|\Delta t|/\tau)^\beta\right) + 2\sigma_n^2.
\label{eq:fullfit}
\end{equation}

We will also test a modification of the above fitting by fixing $\beta=1$ (hence the three-parameter fit) and study the remaining SF parameters.

(2) We will also try another route where we calculate the distribution of $\Delta m$ for $\Delta t<2$ days and use it as the photometric noise estimate 
(because $SF_{\rm true}\rightarrow 0$ as $\Delta t\rightarrow 0$), and we then subtract it in quadrature from Equation~(\ref{eq:sfrmsMcL}):

\begin{eqnarray}
SF_{\rm true}^2(\Delta t) &=& 0.549 \left(IQR^2(\Delta t) - IQR^2(n)\right).
\label{eq:sfrmsMcL2}
\end{eqnarray}

Subsequently, we will fit an SPL SF (Equation~(\ref{eq:sfsinglefit})) for lags in the range $4<\Delta t<365$ days, before the SFs start to flatten. 
Example SF calculations and fits of these methods on the real SDSS data are shown in Figure~\ref{fig:SFfitsexample}.

\begin{figure}
\centering
\includegraphics[width=8.2cm]{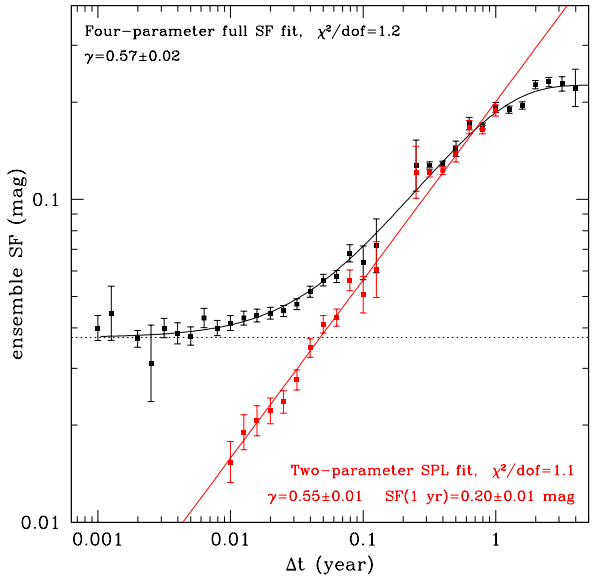}
\caption{Example SF fits for 216 true SDSS AGNs in a bin $z=1.6\pm0.1$ with $M_i=-26.0\pm0.25$ mag. 
The black and red points are calculated from Equations~(\ref{eq:sfrmsMcL}) and (\ref{eq:sfrmsMcL2}), respectively.
The full, four-parameter full SF fit (Equation~(\ref{eq:fullfit})) is marked in black, while
the SPL fit (Equation~(\ref{eq:sfsinglefit})) is marked in red. The dotted horizontal line marks the noise.}
\label{fig:SFfitsexample}
\end{figure}


\subsection{Power Spectral Density}

In this paper, we will not measure the PSD simply
because SFs are much easier to calculate and do not suffer from operational problems like
irregular sampling and data binning. We briefly review the topic because in many AGN variability studies
PSDs are the primary variability measure. We provide a means for their comparison to SFs and their connection to ACFs.

Power spectral density PSD$(\nu)$, that is a function of frequency $\nu$, by definition is
\begin{equation} 
{\rm total~power} \propto \int_{-\infty}^{\infty} {\rm PSD}(\nu) d\nu,
\label{eq:totpow1}
\end{equation}
\noindent 
and PSD$(\nu)$ needs to be in units of power/Hz if the total power is to be expressed in units of power.
Any time series $y(t)$, including AGN variability, can be thought of as a combination of separate periodic signals with frequencies $\nu$. 
Using the Fourier transform, we can decompose such a time series into a series in the frequency domain using (e.g., \citealt{1981ApJS...45....1S,1982ApJ...263..835S,1992nrfa.book.....P})

\begin{equation} 
Y(\nu) = \int_{-\infty}^{+\infty} y(t) e^{-2\pi i \nu t} dt.
\end{equation}

The Parseval's theorem states that the total power in the time domain is equal to the total power in the frequency domain:

\begin{equation} 
{\rm total~power} = \int_{-\infty}^{\infty}|y(t)|^2 dt = \int_{-\infty}^{\infty}|Y(\nu)|^2 d\nu.
\label{eq:totpow2}
\end{equation}
\noindent
The combination of Equations~(\ref{eq:totpow1}) and (\ref{eq:totpow2}) gives
\begin{equation} 
{\rm PSD}(\nu) = |Y(\nu)|^2,
\end{equation}
\noindent
which is simply the square of the Fourier transform of the signal and is expressed in units of mag$^2$~Hz$^{-1}$. 

The spectra of aperiodic signals can be represented as power laws, PSD$(\nu) \propto \nu^{\alpha}$, where
for example $\alpha=0$ is the white noise (i.e. a constant PSD across all $\nu$ values).
The amount of power in any two dex are the same for the pink or flicker noise ($\alpha=-1$), and 
the random walk noise (also Brownian noise) is described by $\alpha=-2$. A typical
AGN PSD is shown in Figure~\ref{fig:vardef}.

A relation of the PSD to ACF can be found via the Fourier transform from 
the Wiener--Khinchin theorem that states 
\begin{equation}
{\rm PSD}(\nu) = \int_{-\infty}^{\infty}{\rm ACF}(\Delta t)e^{-2\pi i \nu \Delta t} \Delta t.
\end{equation}

Because the PSD measures the squared rms (mag$^2$~Hz$^{-1}$) at a frequency $\nu$, 
it can be thought of as ``a reflection'' of an SF (measuring plain rms), where the time lag $\Delta t$ is replaced with
a frequency $\nu$ and the conversion between powers is $\alpha=-4\gamma$.
Please note that some authors present PSD$\times$Hz instead of just PSD and that 
the conversion between powers $\alpha=-2\gamma-1$ from \cite{2009ApJ...696.1241B} is for PSD$\times$Hz
and the square of SF (the ``theoretical'' SF; Equation~(\ref{eq:SFtheoryV})).


\subsection{The DRW model}
\label{sec:DRW}

Although it was noticed over half a century earlier (\citealt{1966SvA....10...15O}), \cite{2009ApJ...698..895K} realized that the stochastic AGN variability, 
in particular the ``red noise'' (PSD $\propto\nu^{-2}$) at short lags and the white noise at long ones, has the same properties 
as the stochastic process called the DRW also known as a
first-order continuous-time autoregressive process (CAR(1)) or Ornstein-Uhlenbeck (OU; \citealt{1930PhRv...36..823U}) process.
Subsequently, it has been shown that aperiodic optical light curves of individual quasars from ground-based studies can be well modeled by DRW 
(\citealt{2009ApJ...698..895K,2010ApJ...708..927K,2010ApJ...721.1014M,2011ApJ...728...26M,2012ApJ...753..106M,2011AJ....141...93B,2011ApJ...735...80Z,2013ApJ...765..106Z,2012ApJ...760...51R}),
while, at the ``very short timescales'' probed by the space-based {\it Kepler} mission and apparently weakly probed by ground-based studies, the PSD shows a departure from DRW (\citealt{2011ApJ...743L..12M,2015MNRAS.451.4328K}; we will discuss this in Section~\ref{sec:discussion}). 
\cite{2014ApJ...788...33K} provide a more general approach to modeling unevenly sampled stochastic light curves using the CARMA models.
DRW is the simplest of the CARMA($p,q$) models, with $p=1$ and $q=0$, and as a matter of fact only DRW can be identified with the physical processes such as Brownian motions
or the Wiener process having a natural interpretation of the model parameters.

A quasar light curve can be described (and fitted) with the DRW model, having just two model parameters: the damping timescale $\tau$ (the same decorrelation timescale as in SFs) 
and the modified variability amplitude $\hat{\sigma}=\sigma\sqrt{2/\tau}$ (or equivalently SF$_\infty=\hat{\sigma}\sqrt{\tau}=\sqrt{2}\sigma$; \citealt{2010ApJ...721.1014M}). 
The ACF for DRW has the form

\begin{equation}
{\rm ACF}(\Delta t) = e^{-|\Delta t|/\tau}
\end{equation}
\noindent
and can be generalized as the PE (e.g., \citealt{2013ApJ...765..106Z})

\begin{equation}
{\rm ACF}(\Delta t) = e^{-(|\Delta t|/\tau)^\beta},
\label{eq:ACFPE}
\end{equation}
\noindent
for $0<\beta<2$ and $\tau>0$, where $\beta=1$ corresponds to DRW (Figure~\ref{fig:PE}). The parameter $\beta$ is responsible not only for the shape of
 the ACF but also for the SF and PSD slopes. Small values of $\beta$ drive the correlation to be weaker, which in turn produces 
flatter SFs (Figure~\ref{fig:PE}). In opposition, large values of $\beta$ make the correlation stronger, which causes the SF to steepen.

In order to understand both the SFs and the recovered DRW parameters from the real SDSS data, we will perform simulations of light curves using DRW.
We generate 100-year-long light curves (to make sure the signal covariance between points is zero, also known as the red noise leakage) using the prescription of \cite{2010ApJ...708..927K}. 
The light curves will be later cut into shorter time baselines or have the cadence reduced or include seasonal gaps, nominally degraded to a typical ground-based survey (Figure~\ref{fig:exampLCS}). 
In short, the chain is initiated by $s_1=G(\sigma^2)$, where $G(x^2)$ is a Gaussian deviate of
dispersion $x$. The subsequent light curve points come from
\begin{equation}
s_{i+1}=s_i e^{-\Delta t/\tau} + G\left[\sigma^2\left(1-e^{-2\Delta t/\tau}\right)\right],
\label{eqn:nextpoint}
\end{equation}
where $\Delta t = t_{i+1}-t_i$ is the time interval.
The observed light curve is obtained from $y_i=s_i+G(n_i^2)$, where $n_i$
is the observational noise. Simulations of generic light curves for any covariance functions
are explained in, e.g., \cite{2011ApJ...735...80Z} and \cite{2016MNRAS.459.2787K}.

\begin{figure}[h]
\centering
\includegraphics[width=8.2cm]{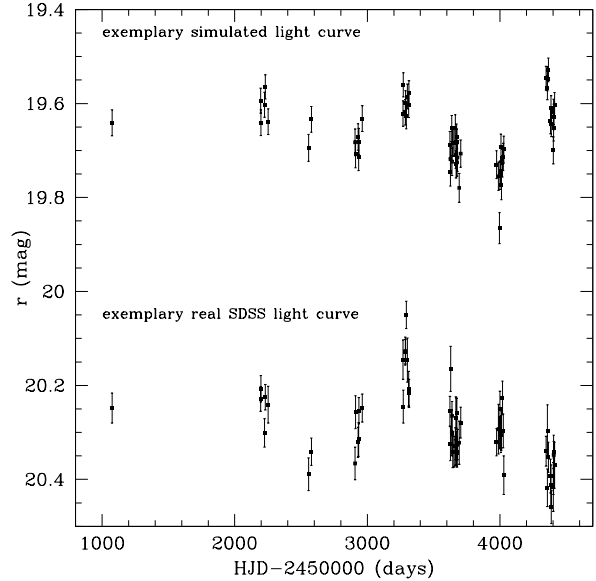}
\caption{Example light curves for a simulated (top) and true SDSS (bottom) AGN variability.}
\label{fig:exampLCS}
\end{figure}


\section{Revisiting AGN Variability in SDSS}
\label{sec:data}

We will now analyze a sample of $\sim$9000 SDSS AGNs, considered by \cite{2010ApJ...721.1014M}, to study their variability through the SF analysis and DRW light curve modeling. 
Of the five SDSS filters, the $r$-band has the highest throughput (\citealt{1996AJ....111.1748F}) and we will be primarily considering this filter.
The light curves come from data reductions by \cite{2007AJ....134..973I}. The AGN identifications are taken from the SDSS Data Release 5/7 (\citealt{2009ApJS..182..543A,2007AJ....134..102S}),
while the basic physical parameters such as the absolute magnitudes come from \cite{2007AJ....134..102S}, and the black hole masses and bolometric luminosities are from
\cite{2008ApJ...680..169S}. Whenever necessary, we calculate our own K-corrections using the SDSS filter throughputs and the mean AGN spectrum from \cite{2001AJ....122..549V}.
We have used the standard cosmological LCDM model with $(H_0, \Omega_M, \Omega_{\rm vac}, \Omega_k)=(70~{\rm{km~s^{-1} Mpc^{-1}}}, 0.3, 0.7, 0.0)$.

\subsection{Data Analysis}

To subtract the noise term correctly (or to fit it correctly) or for it to have well-behaved properties,
we will be considering a ``subensemble'' variability of AGNs in the narrow redshift--absolute magnitude bins of the size 0.2--0.5 mag, with more than 10 AGNs per bin (median 50 AGNs).
\cite{2010MNRAS.404..931E} provide an overview of an SF analysis and notes that breaks and wiggles at long timescales in a single SF can occur even 
in a process not having such an intrinsic break timescale.
(We observe an analogous behavior, but we also find that averaging a large number of SFs causes such breaks and wiggles to vanish, which in a statistical sense means that 
they are no longer outliers from the input SF (see Figure~\ref{fig:SFexplain2}).)

Each AGN light curve is converted into a file with the time differences between epochs and the corresponding differences of magnitudes for these epochs ($dt_i$, $dm_i$), 
where the time differences are corrected for redshift. In each redshift--absolute magnitude bin, we combine all AGN files and  
calculate distributions of $\Delta m$ in logarithmic bins of the time difference $\Delta t$ of width $0.1$ dex and use Equation~(\ref{eq:sfrmsMcL})
to estimate the rms variability at each $\Delta t$. Please note this is now the SF that includes the noise term. 

We estimate the SF error bars in each $\Delta t$ bin by simulation means. 
(1) We have tested the bootstrap method. (2) We have tested the dispersions of IQRs of randomly selected subsamples of the original distribution. 
(3) We generated a large number of pure Gaussian distributions with $N$ data points and empirically found that the uncertainty of the SF measurement in a bin, 
calculated as $SF=0.741\times IQR$, is simply $\Delta SF=1.17 \times SF \times N^{-1/2}$ and equals to the dispersion of the SF for the various Gaussian distribution realizations. 
Because in every $\Delta t$ bin we combine the magnitude differences from a number of different AGN light curve realizations 
from a narrow but nonzero spread of magnitudes, we find that the bin-to-bin SF variations from the three methods are much larger than these estimated uncertainties.
The optimal noise estimate is different, and we adopt $\Delta SF=4 \times N^{-1/2}$ as our SF uncertainty
in order to obtain $\chi^2/{\rm dof}\approx1$ across the full redshift--absolute magnitude plane for the SF fits to the simulated data.
Most importantly, such an approach returns correctly measured SF parameters (see the simulations in Figures~\ref{fig:simulPE}--\ref{fig:simulSPL}).

\begin{figure*}
\centering
\includegraphics[width=6.8cm]{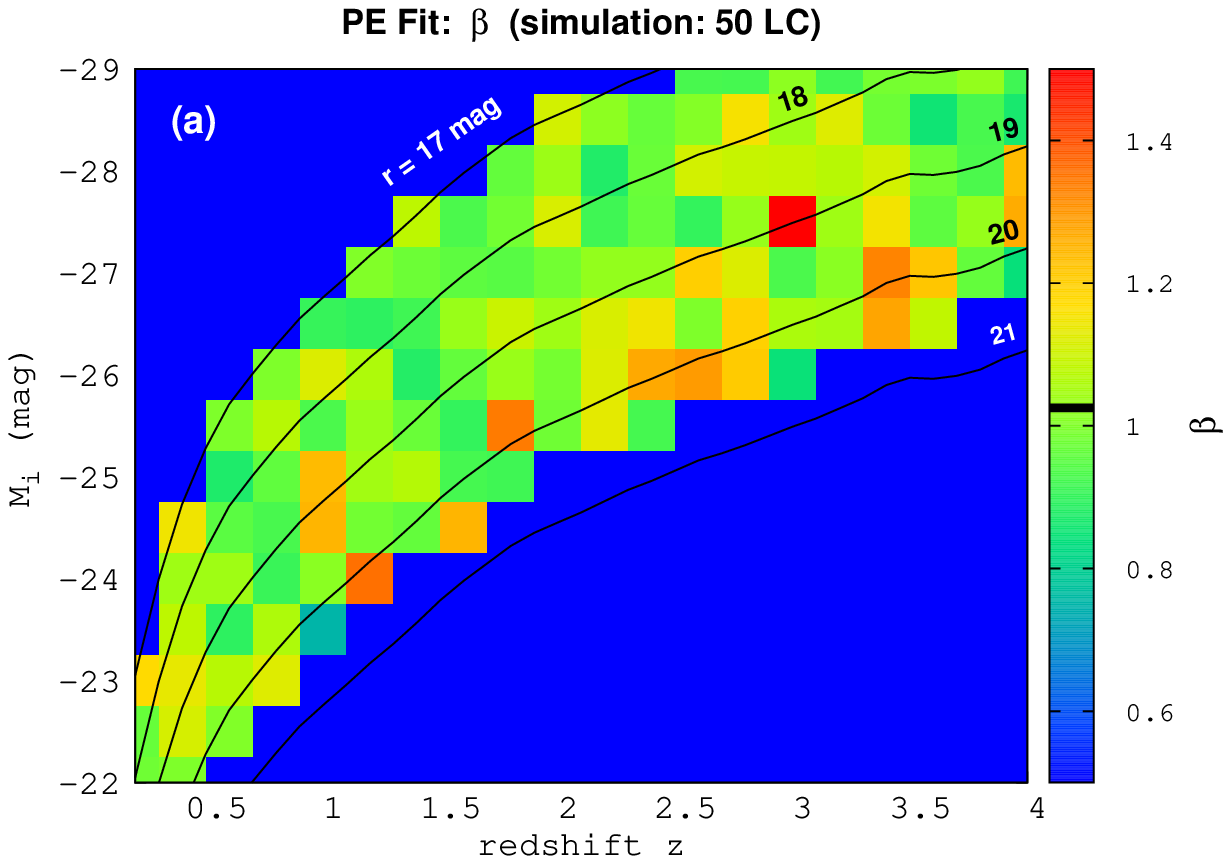} \hspace{0.1cm}
\includegraphics[width=6.8cm]{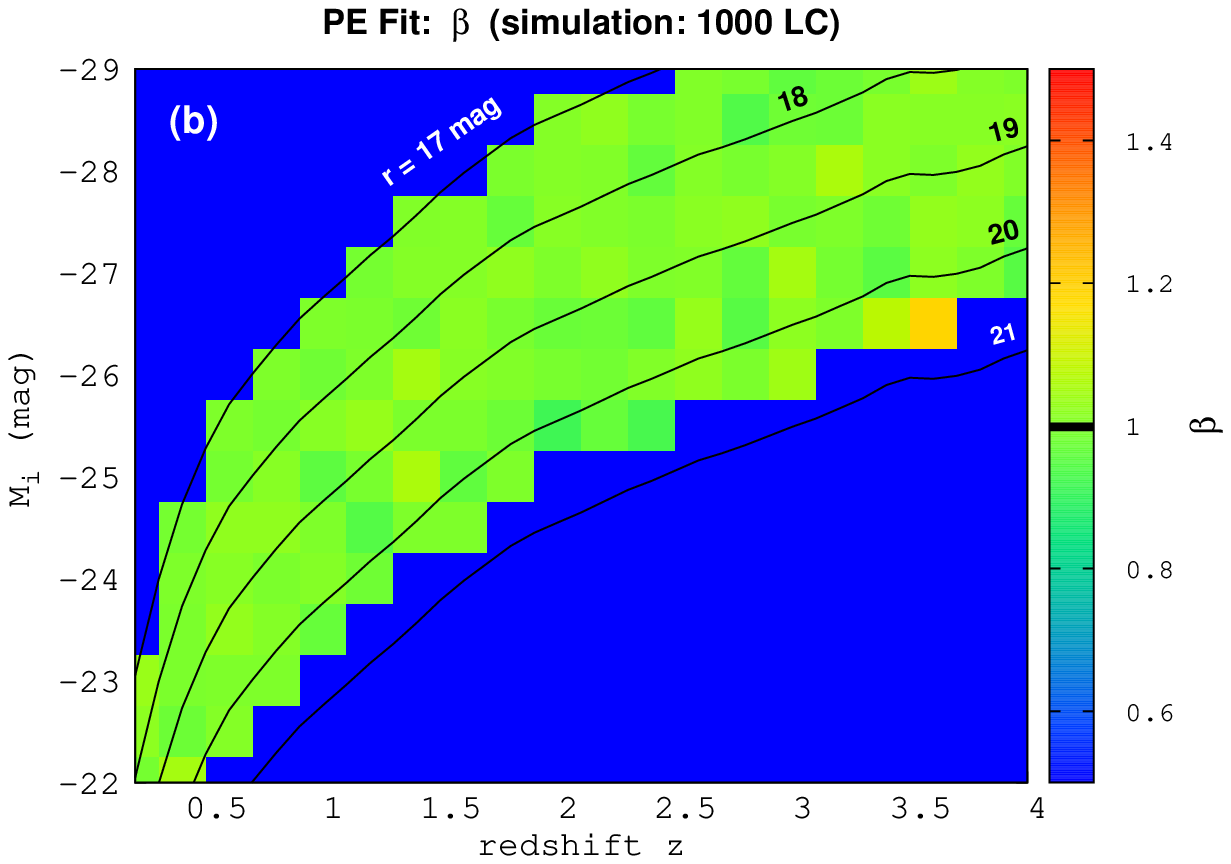}\\
\vspace{0.3cm}
\includegraphics[width=6.8cm]{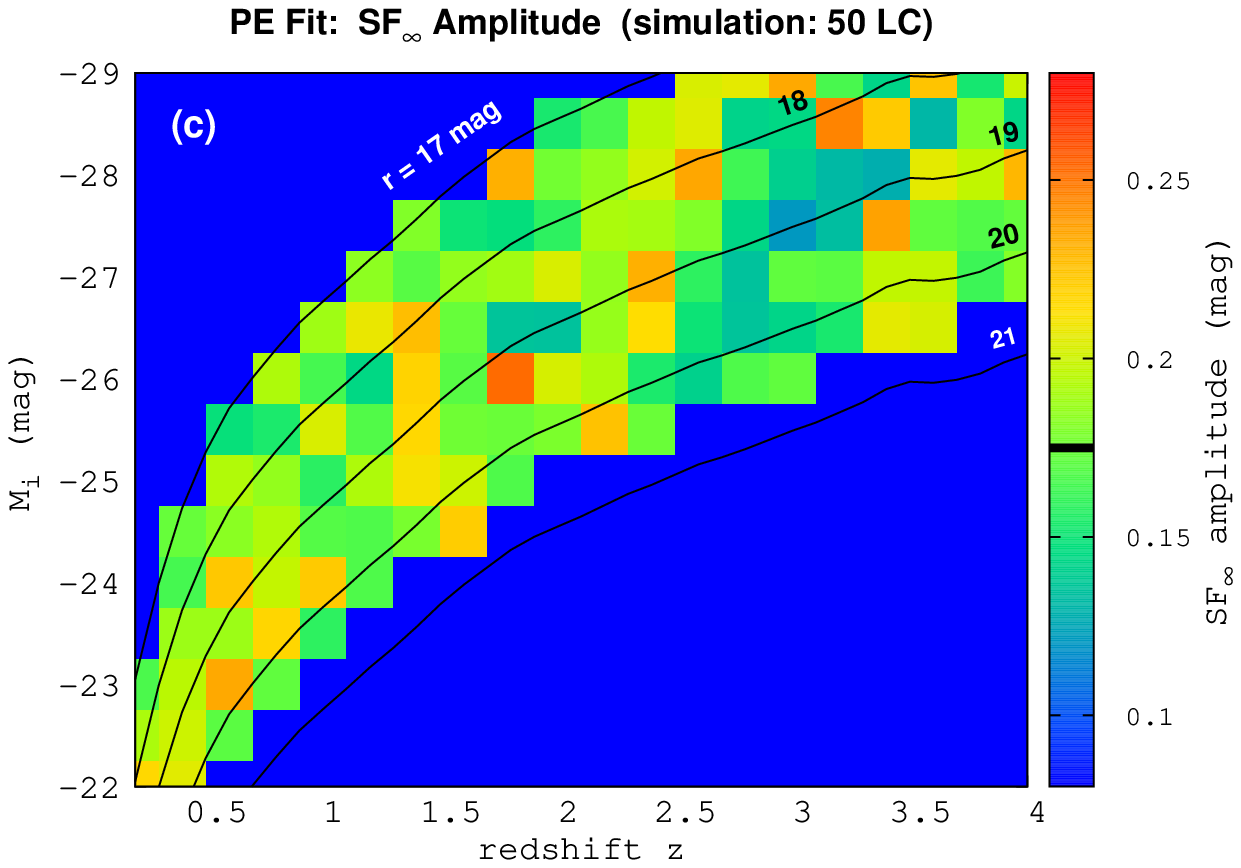} \hspace{0.1cm}
\includegraphics[width=6.8cm]{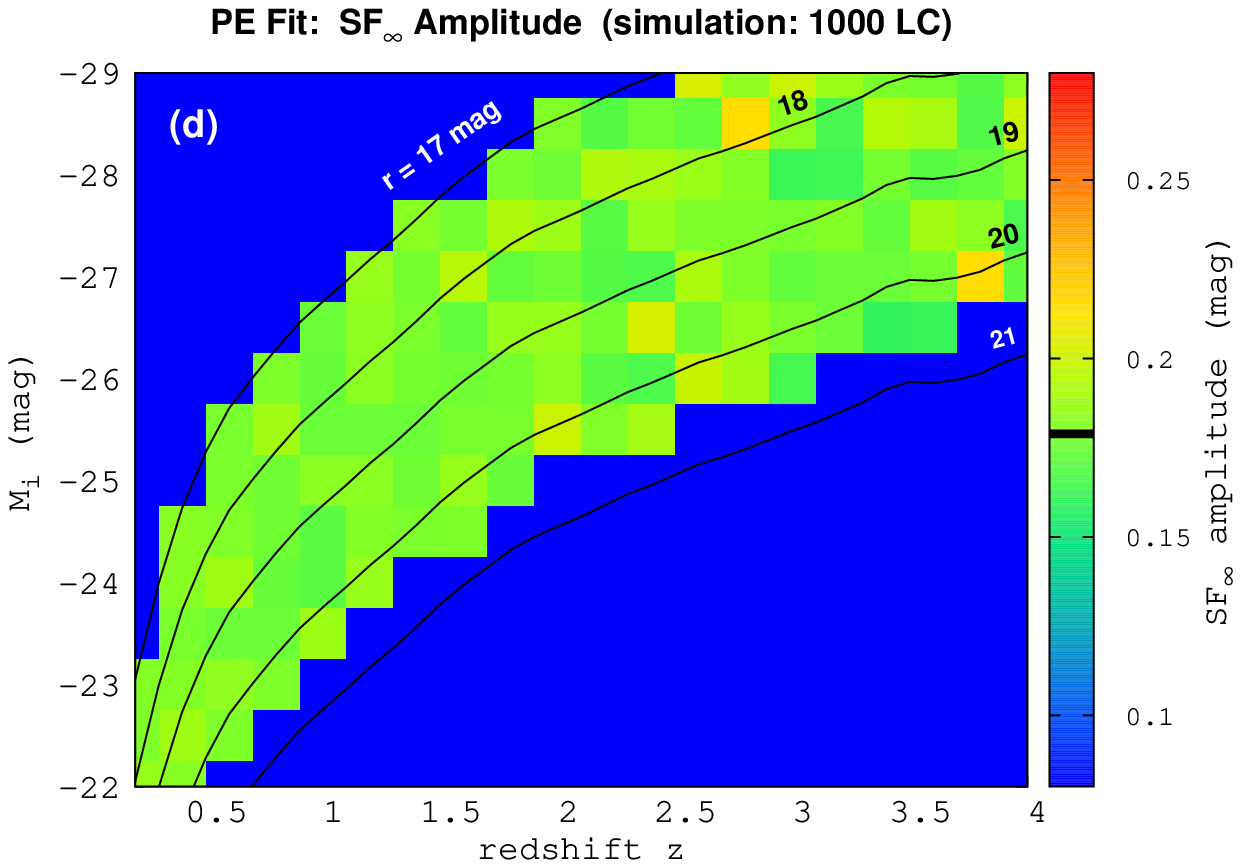}\\
\vspace{0.3cm}
\includegraphics[width=6.8cm]{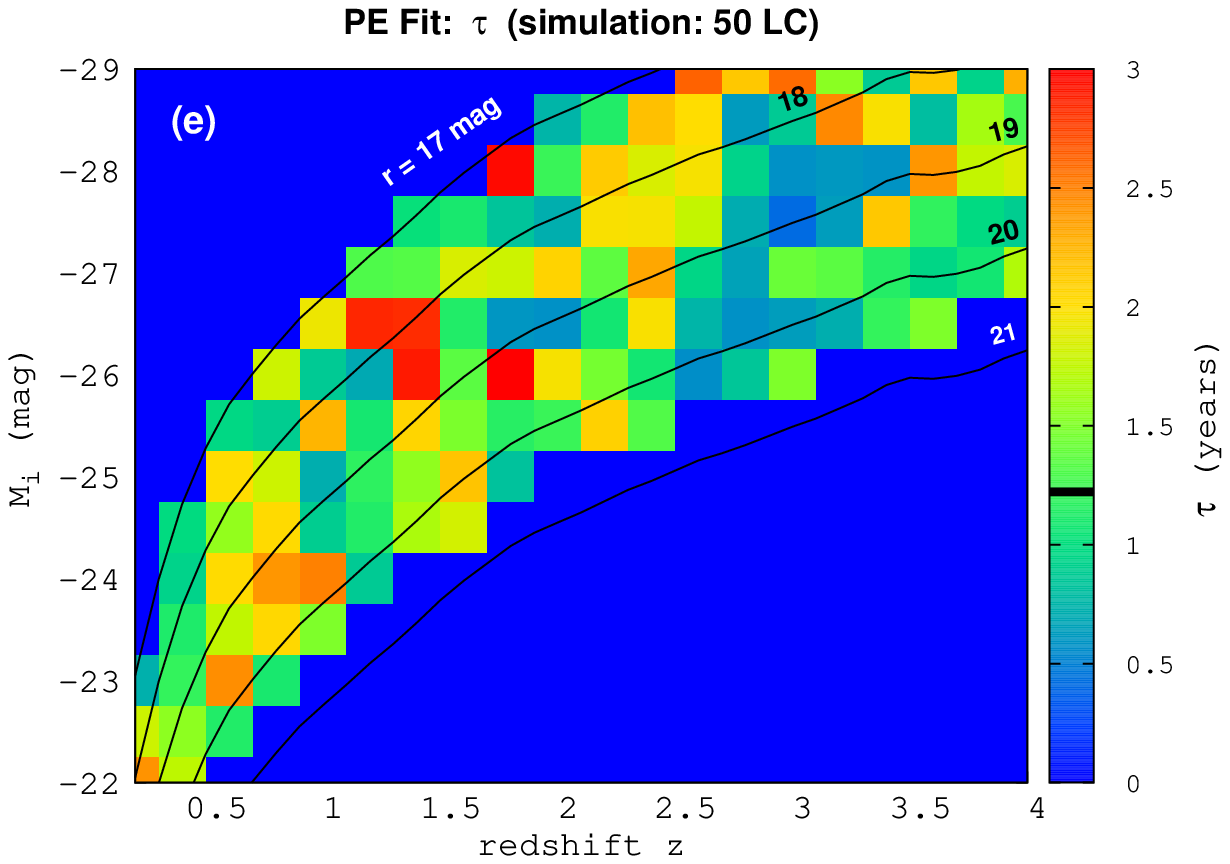} \hspace{0.1cm}
\includegraphics[width=6.8cm]{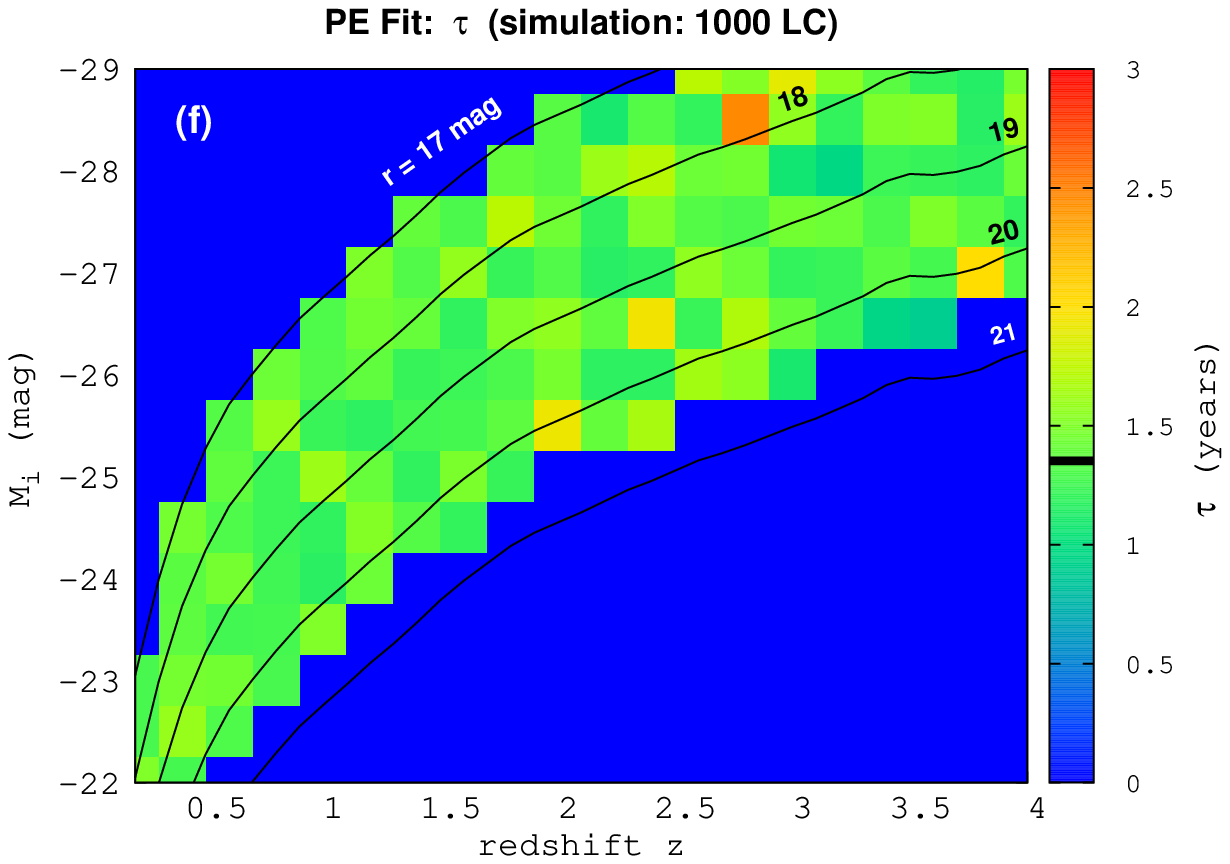}\\
\vspace{0.3cm}
\includegraphics[width=6.8cm]{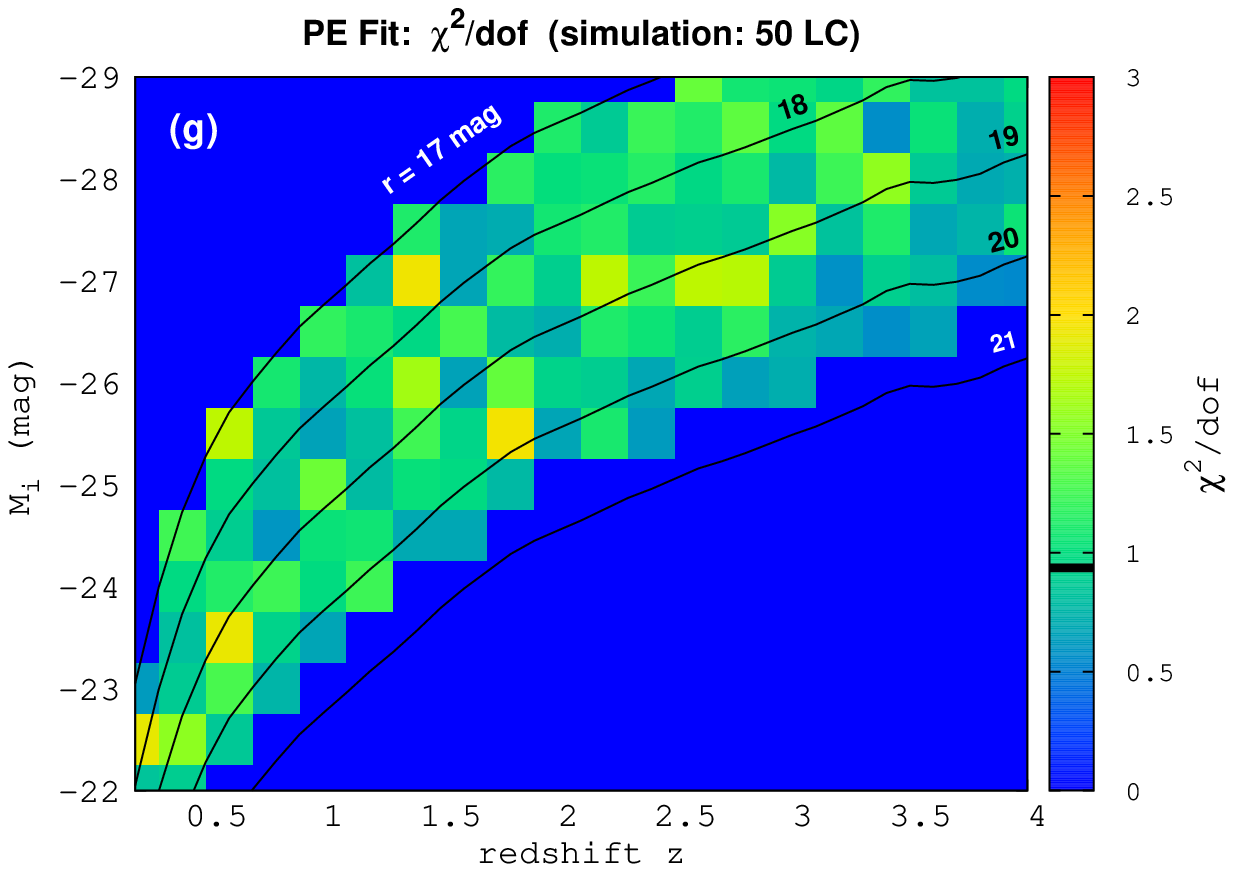} \hspace{0.1cm}
\includegraphics[width=6.8cm]{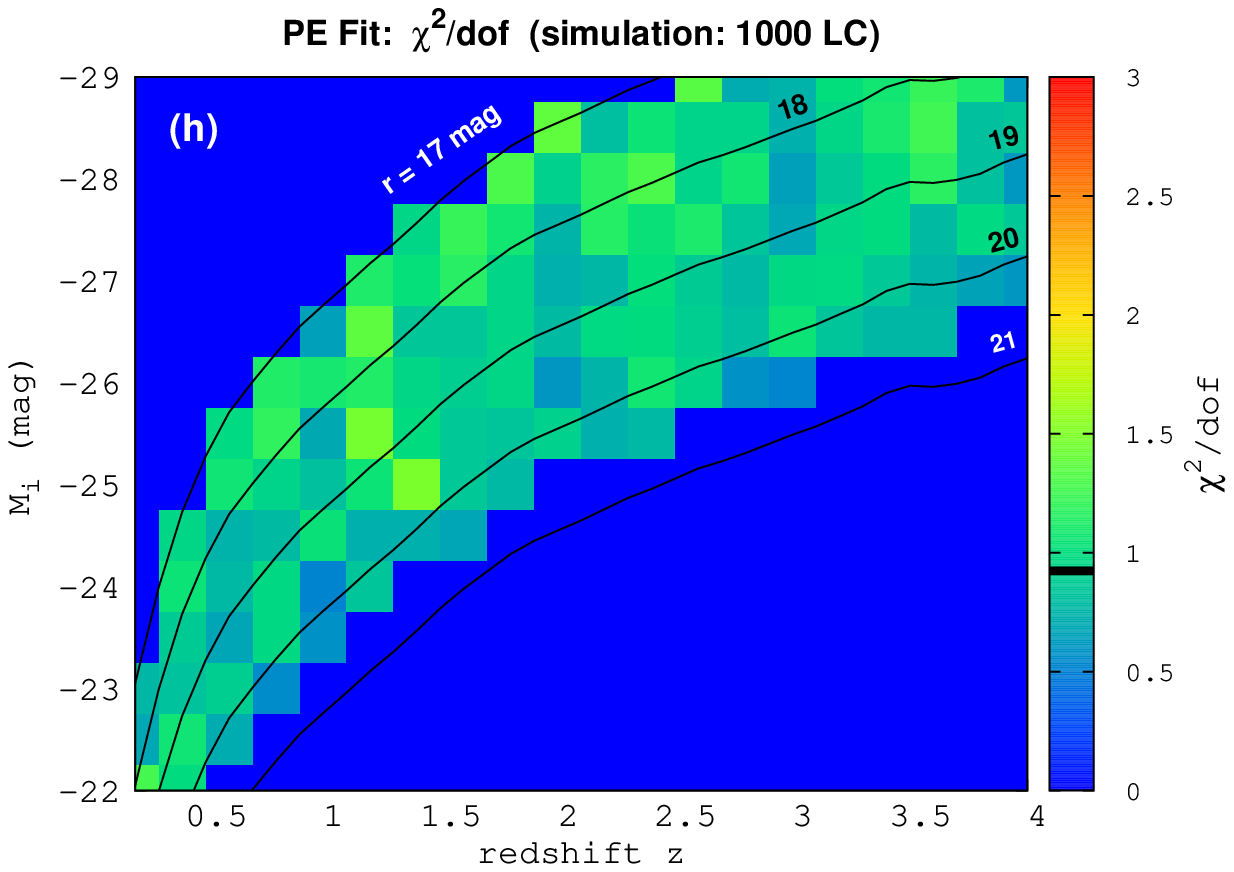}
\caption{The recovered SF parameters from simulations of 50 (left column) and 1000 (right) AGN light curves per redshift--absolute magnitude bin. 
The light curves were simulated as the DRW stochastic process ($\beta=1.0$) with the input parameters $\tau=500$ days (1.37 years) and $SF_\infty=0.18$ mag.
All panels show the same redshift--absolute magnitude ranges and also the lines (black) for the constant observed $r$-band magnitude. 
Each panel is complemented by a dedicated color scale on its right that spans an appropriate range of the parameter space.
The median values (weighted with the AGN number) are marked on the color bars with the thick black line.
The light curves were fit with the full, four-parameter SF function that measures
the power $\beta$ of the power exponential (PE) covariance matrix of the signal. The recovered parameters are nearly identical to the input ones and hence are
unbiased functions of the absolute magnitude and redshift.}
\label{fig:simulPE}
\end{figure*}


\begin{figure*}
\centering
\includegraphics[width=6.8cm]{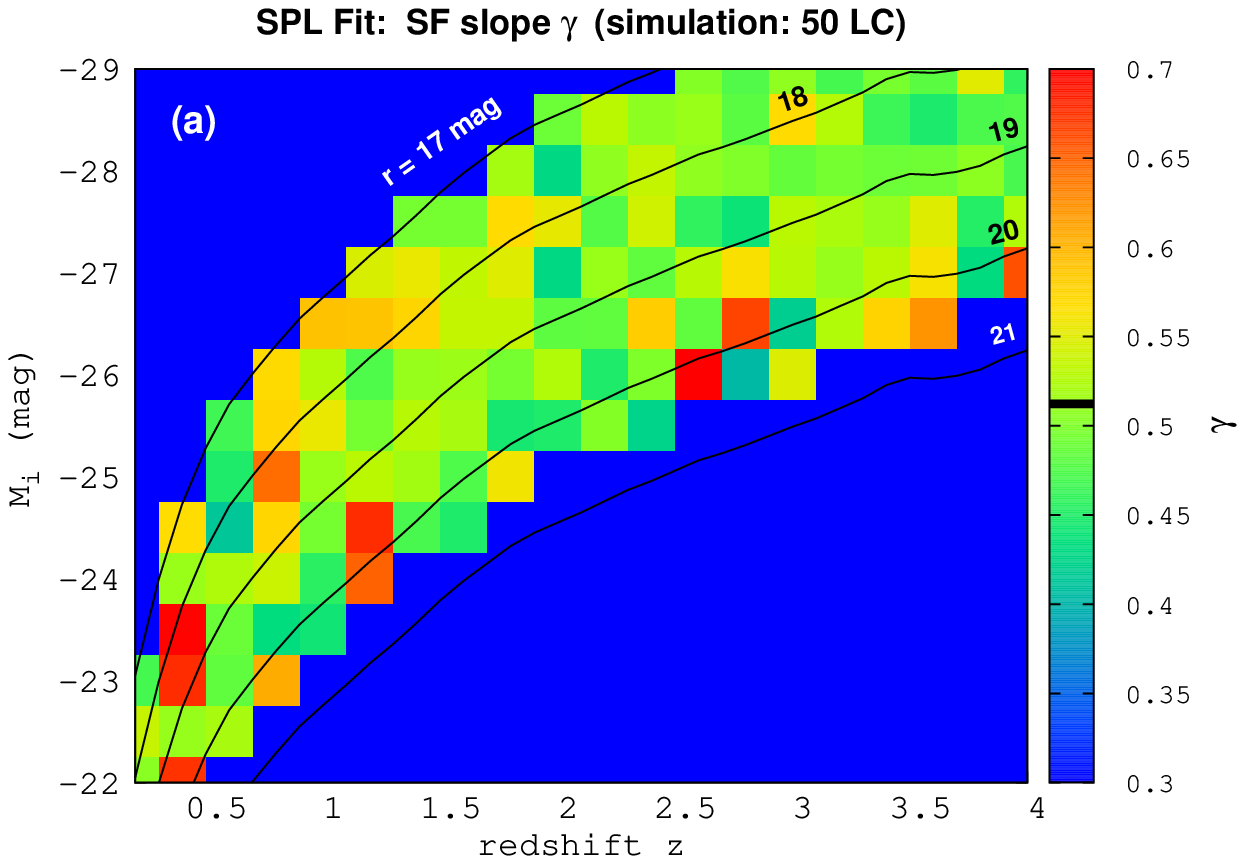} \hspace{0.1cm}
\includegraphics[width=6.8cm]{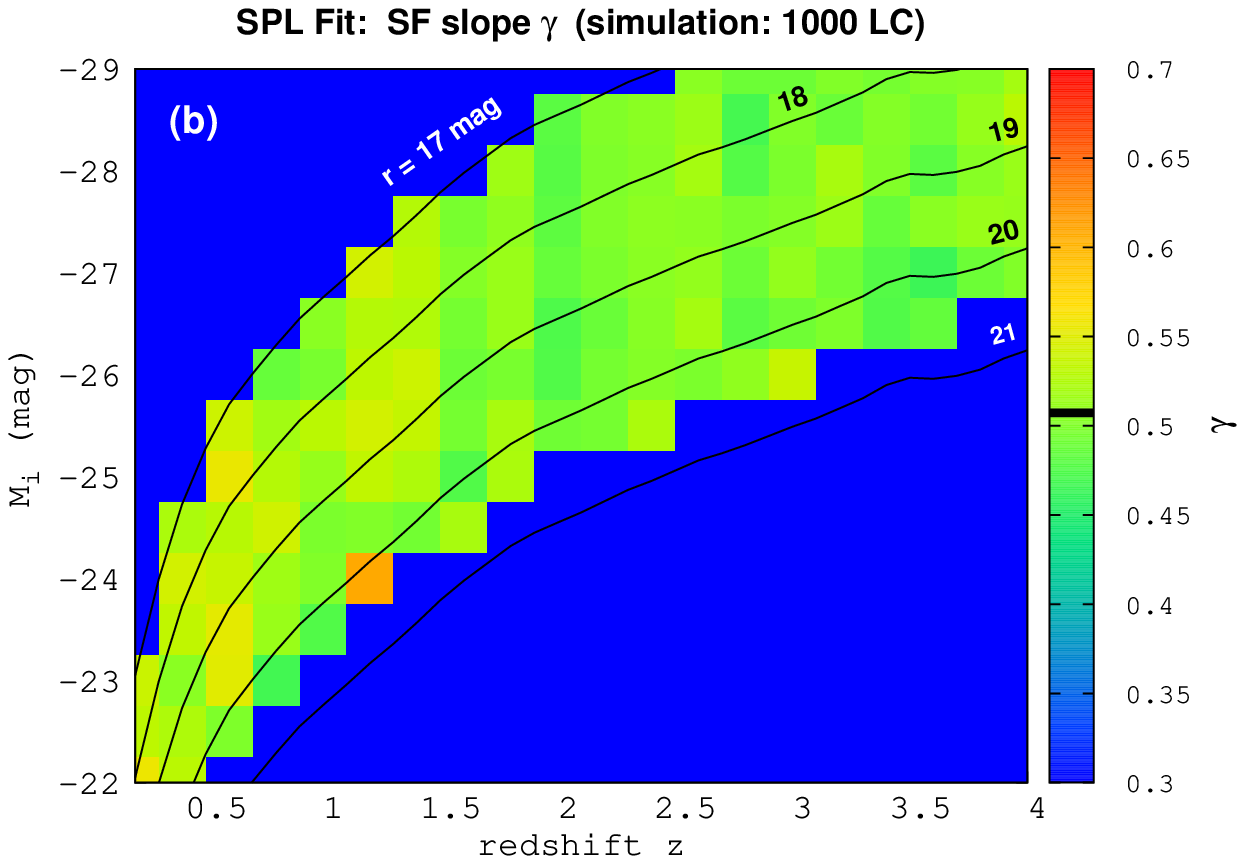}\\
\vspace{0.3cm}
\includegraphics[width=6.8cm]{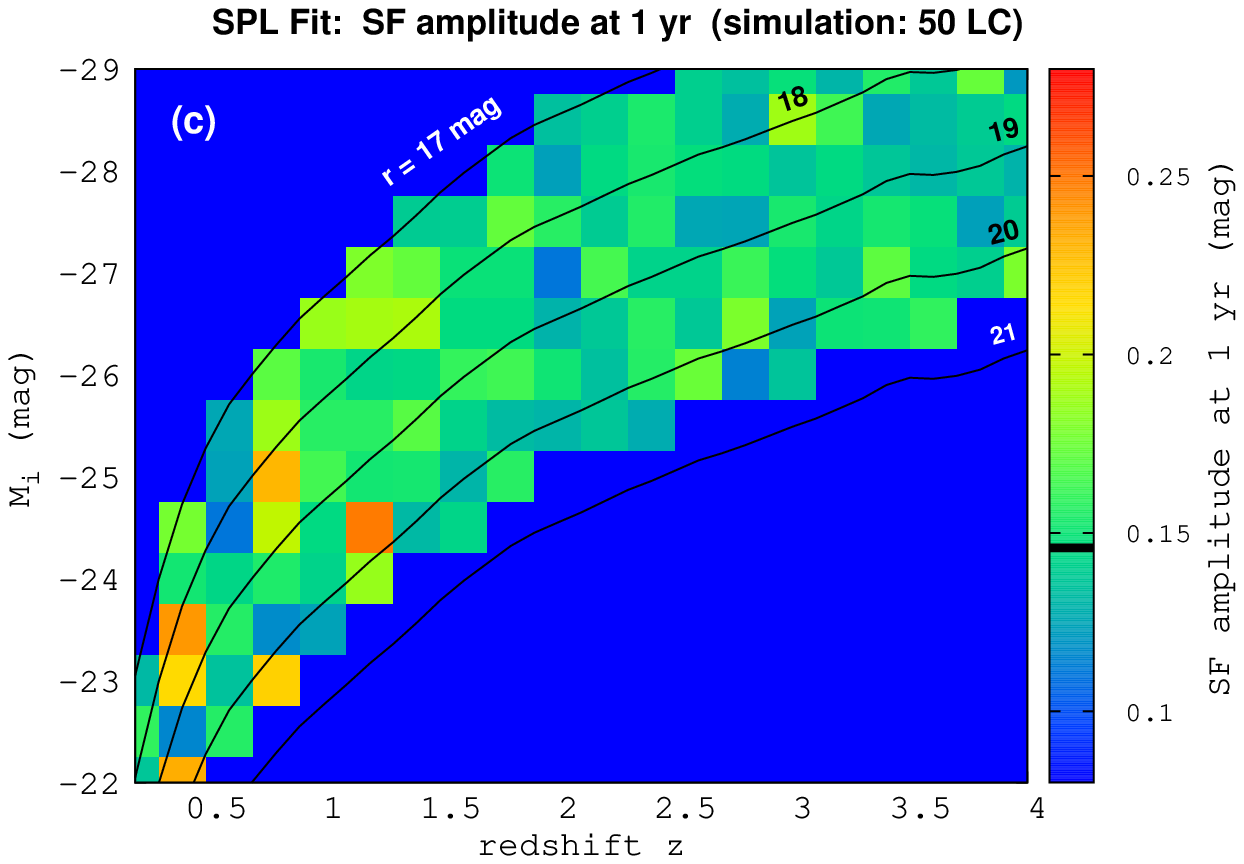} \hspace{0.1cm}
\includegraphics[width=6.8cm]{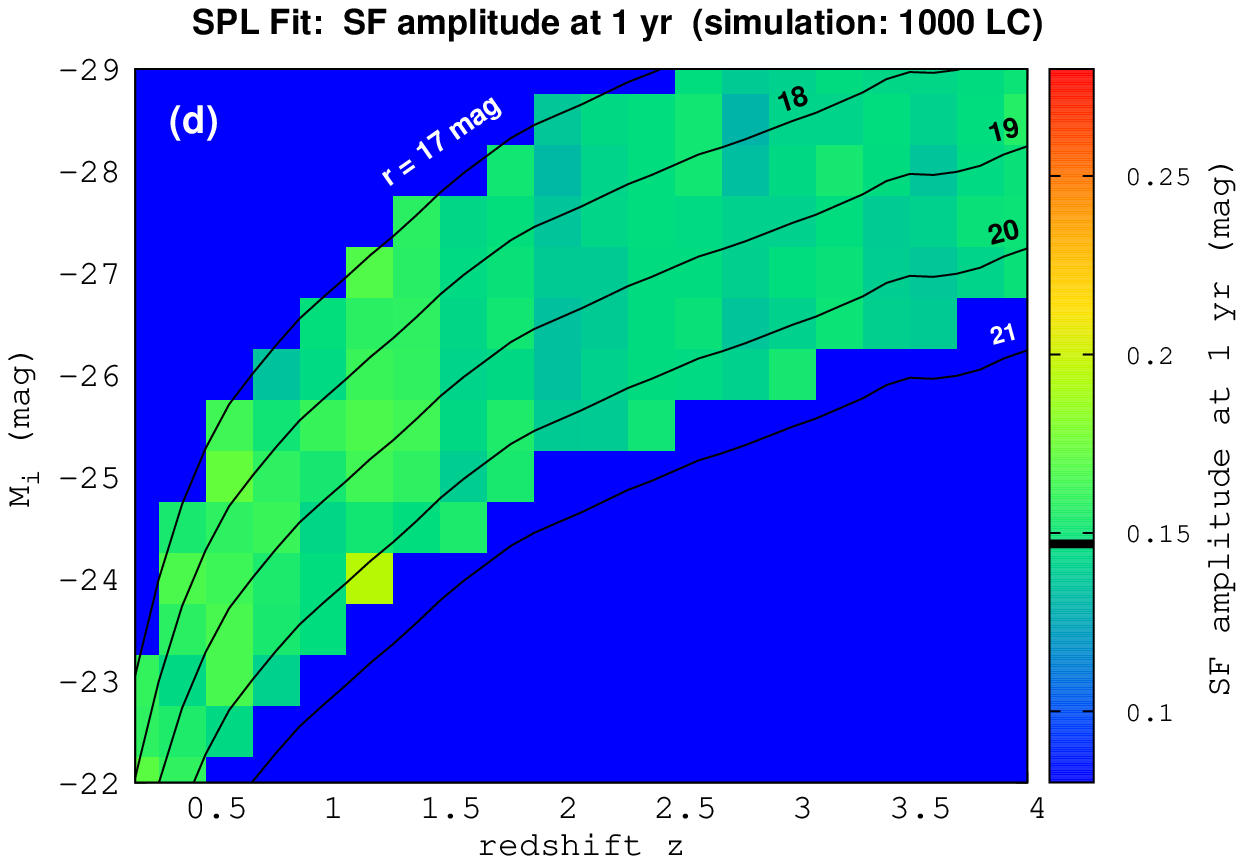}\\
\vspace{0.3cm}
\includegraphics[width=6.8cm]{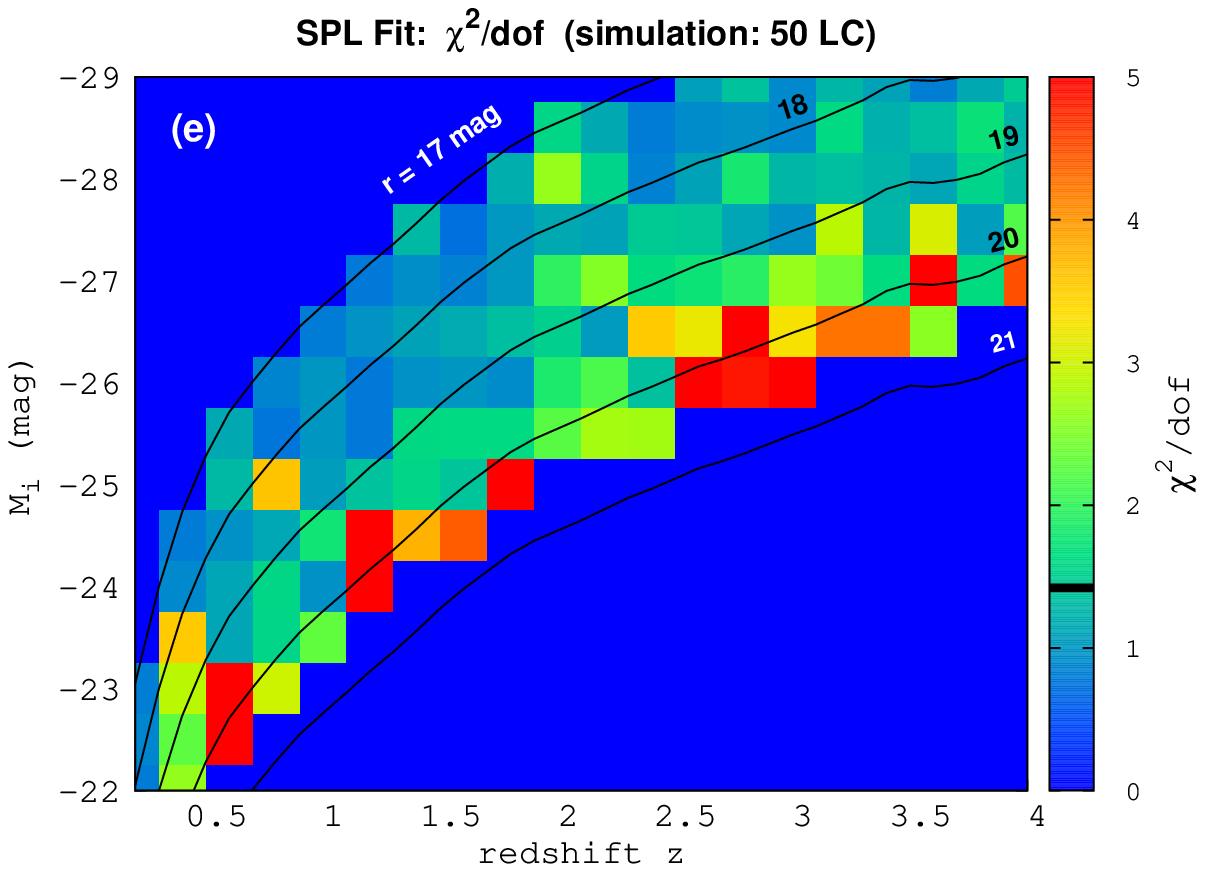} \hspace{0.1cm}
\includegraphics[width=6.8cm]{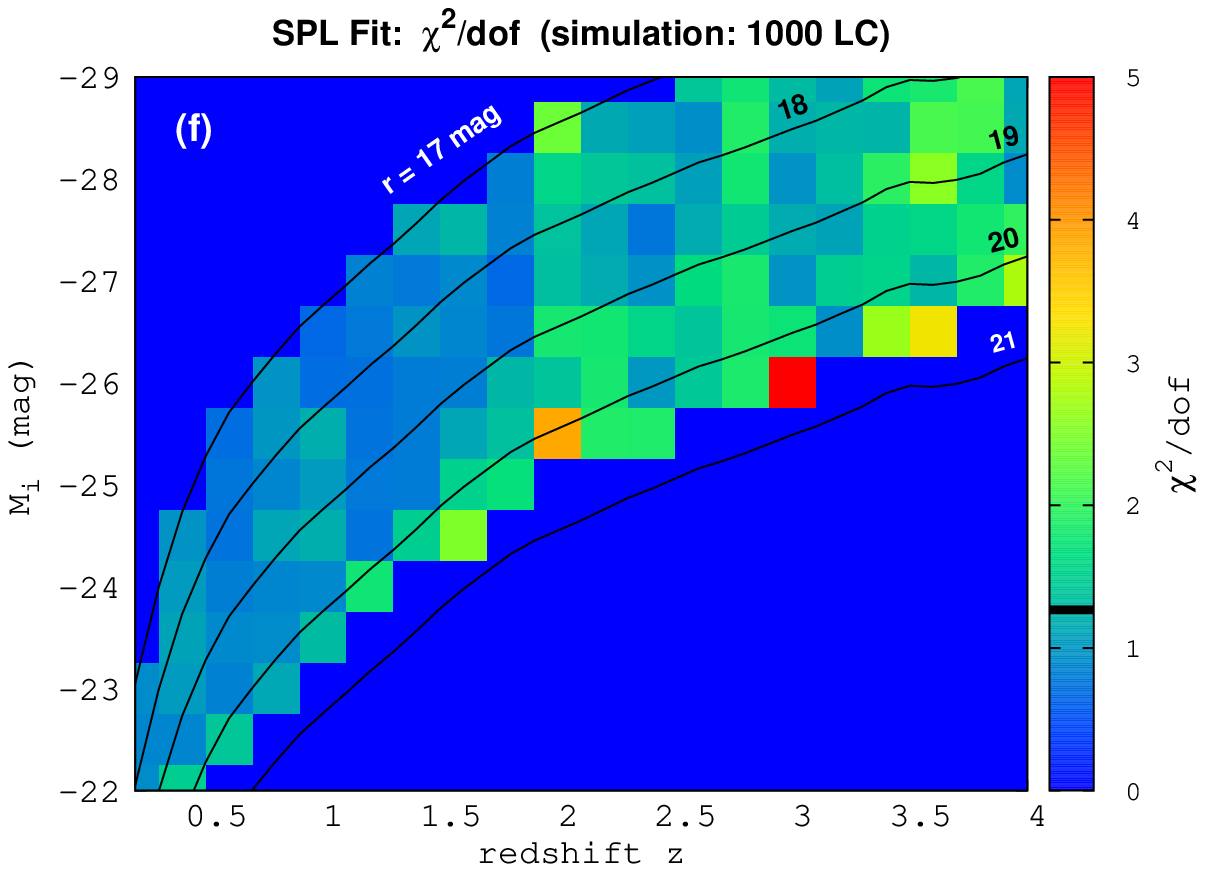} 
\caption{The recovered SF parameters from simulations of 50 (left column) and 1000 (right) AGN light curves per redshift--absolute magnitude bin. 
The light curves were simulated as the DRW stochastic process ($\gamma=0.5$) with the input parameters $\tau=500$ days and $SF_\infty=0.18$ mag (SF at one year is 0.15 mag).
All panels show the same redshift--absolute magnitude ranges and also the lines (black) of the constant observed $r$-band magnitude. 
Each panel is complemented by a dedicated color scale on its right that spans an appropriate range of the parameter space.
The median values (weighted with the AGN number) are marked on the color bars with the thick black line.
The light curves were fit with a single power law (SPL) SF function in the range  $4<\Delta t<365$ days. 
There is a small bias in the recovered SPL slope $\gamma=0.52$, as compared to the input 0.5, that slightly increases toward lower redshifts.
The amplitude shows a similar behavior, but it is smaller at higher redshifts than the input value.}
\label{fig:simulSPL}
\end{figure*}

We will be considering two fitting methods now (described in Section~\ref{sec:sfcalcfit}):

(1) We will fit the SF using the full, four-parameter SF model with Equation~(\ref{eq:fullfit}) and employ the {\sc MINUIT} minimization routines\footnote{\tt www.cern.ch/minuit}.
The SDSS S82 light curves are generally short (in the rest frame), weakly probing the SF turnover, 
so in the minimization procedure we have added a weak prior on the timescale, $\Delta \chi^2=(\tau-1.5~{\rm yr})^2$, 
to prevent it from going to infinity, where $\sim$1.5 year is the typical AGN timescale found from DRW modeling in \cite{2010ApJ...721.1014M}.
In the majority of the bins the timescale is recovered correctly without this additional prior.

Next, we make a small modification to method (1), where we fix $\beta=1$. It is clear we will not measure the SF slope, but we are going to study the remaining SF parameters.
In particular, from method (1) (as we will show later) we see that the SF slope slightly steepens with the increasing luminosity, but the SF decorrelation timescale stays constant. 
From  direct light curve modeling with DRW, which has the fixed $\beta=1$, we see an increase of the timescale with the increasing luminosity.
This method will be primarily used to check if the DRW timescale increases with the increasing $\beta$ (and the answer ahead is yes).

(2) In this method, we calculate distributions of $\Delta m$ for $\Delta t<2$ days and use it as a noise estimate 
that we subtract in quadrature from the SFs in method (1); see Equation~(\ref{eq:sfrmsMcL2}).
This is a noise-free, true SF that can be fitted with an SPL, Equation~(\ref{eq:sfsinglefit}), in the timescale range $4<\Delta t<365$ days, that is, 
before the SF starts to flatten in the vicinity of one year. Because this is a simple line fitting, we use a least-squares $\chi^2$ minimization (as explained by, e.g., \citealt{2003astro.ph.10577G}).

Example calculated SFs and best fits obtained with methods (1) and (2) are shown in Figures~\ref{fig:SFexplain2}--\ref{fig:SFfitsexample}.

\subsection{Simulations of the SDSS data}

Prior to estimating the variability parameters from the real data, we have to understand any possible biases or systematic effects present in the methods used to measure the SFs. 
We do this by simulating artificial AGN light curves in two samples with 50 and 1000 AGNs per bin. The first one will show us approximately the expected scatter between bins for the real data, while the latter will enable us to track down any low-level systematics and biases. We simulate 100-year-long light curves with a cadence of four days. From such a long time series, we
pick only these epochs matching the real ones (typically 60 epochs) that occur at least 40 years after the simulation starts. This is to make sure all correlations
at long $\Delta t$ are included (and there is no red noise leakage). The exact procedure for generating the light curves is detailed in Section~\ref{sec:DRW}. They are simulated with
the DRW model, so the expected SF slope for $\Delta t \ll \tau$ is $\gamma\equiv0.5$. The input parameters are picked to be  
$\tau=500$ days and $SF_\infty=0.18$ mag (in fact, $\tau=500(1+z)$ days, and then the simulated light curves are corrected for the $(1+z)$ term).

The results of the simulations are presented in Figures~\ref{fig:simulPE}--\ref{fig:simulSPL}, where the
left (right) column presents the subensemble variability results for 50 (1000) AGNs light curves per bin.
We will be interested in any biases or systematics between the measured output variability parameters and the input ones that are known.
In Figures~\ref{fig:simulPE}--\ref{fig:simulSPL}, each panel is complemented by a dedicated color bar spanning an adequate parameter range.
To estimate the median value across the redshift--absolute magnitude plane, we create a distribution of values by counting the measured parameter in every bin a number of times equal to the 
number of AGNs in that given bin. The median value from such a distribution is measured and marked as a thick black line in the color bars of Figures~\ref{fig:simulPE}--\ref{fig:simulSPL}.

In Figure~\ref{fig:simulPE}, we see that, for the simulations with both 50 and a 1000 light curves per bin, the recovered parameters from the full SF fitting are stable and unbiased functions of
redshift and the absolute magnitude. The median $\beta=1.016$ ($\gamma=0.5008$) is nearly identical to the input value $\beta_{\rm input}=1.0$,
with median $SF_\infty=0.175$ mag and $0.179$ mag for the 50 and 1000 light curves, respectively, while $SF_{\infty \rm ~input}=0.18$ mag.
The input timescale is $\tau_{\rm input}=500$ days (1.37 year), and the measured values are $\tau=1.22$ and $\tau=1.35$ years for the 50 and 1000 light curves.

In Figure~\ref{fig:simulSPL}, we present the best-fit parameters obtained from an SPL SF fitting.
We observe very weak trends for $\gamma$ and the SF amplitude at one year with redshift.
The distribution of recovered $\gamma$ peaks at 0.52 for bins with 50 light curves, i.e., slightly higher than the DRW input of $\beta_{\rm input}=1.0$. 
We will be later correcting the $\gamma$ obtained from the real data for this small bias.

\subsection{The real SDSS data}

In panels (a) and (b) of Figures~\ref{fig:realPE} and \ref{fig:realSPL}, we present the mean values (in a bin) of the black hole mass and the Eddington ratio, respectively.
It is clear that both these parameters increase with the increasing AGN luminosity. In panels (c) and (d) of Figure~\ref{fig:realPE}, we present the measured values of 
the power $\beta$ of the PE covariance matrix and the amplitude at long timescales ($\Delta t\rightarrow \infty$), $SF_\infty$, respectively.
It appears that $\beta$ slightly increases ($SF_\infty$ decreases) with increasing luminosity. From the simulations, we know these parameters should be unbiased 
functions of redshift and luminosity. The measured decorrelation timescales seem to be constant across the redshift--absolute magnitude plane, panel (e) of
Figure~\ref{fig:realPE}, and the goodness of fits seems to be also reasonable, panel (f). We also model SFs with a three-parameter fit (with fixed $\beta=1$).
In panels (g) and (h), we show the recovered timescales with and without the weak prior on the timescale ($\Delta \chi^2=(\tau-1.5~{\rm yr})^2$), respectively.
We see that by fixing $\beta=1$, the steepening of $\beta$ in the four-parameter fit is now replaced by the increasing timescale, in a similar fashion
to what is observed for direct light curve modeling with DRW (Figure~\ref{fig:realSPL}, panel (g)).

In panels (c) and (d) of Figure~\ref{fig:realSPL}, we present the SF slope $\gamma$ and the amplitude at one year, measured from the SPL fitting
in the timescale range $4<\Delta t<365$ days. We see a similar dependence of $\gamma$ with the luminosity as in the case of $\beta$ in panel (c) of Figure~\ref{fig:realPE};
that is, $\gamma$ increases when the luminosity increases. Also similarly to $SF_\infty$ in panel (d) of Figure~\ref{fig:realPE}, 
the amplitude at one year decreases with the increasing luminosity in panel (d) of Figure~\ref{fig:realSPL}. Panels (e) and (f) present 
the number of SDSS Stripe 82 AGNs per redshift--absolute magnitude bin and the goodness of fits, respectively.
We also modeled the SDSS light curves directly using the DRW model (see \citealt{2010ApJ...708..927K} for details). 
In panels (g) and (h), we present the mean values of the DRW parameters $\tau$ and $\hat{\sigma}$, respectively.
It is clear that the DRW timescale $\tau$ shows a similar dependence on redshift and the absolute magnitude as parameters $\beta$ and $\gamma$,
while the $\hat{\sigma}$ is akin to the amplitude at one year and $SF_\infty$.

We also perform simulations of 1000 DRW light curves per bin to measure the biases in the parameters $\tau$ and $\hat{\sigma}$
in the redshift--absolute magnitude plane (Figure~\ref{fig:simulDRW}). From panel (a), it is very clear that 
the recovered timescales are typically underestimated as compared to the input ones, 
and this ratio decreases with the increasing ratio of the input timescale to the experiment length.
In panel (b), we show the parameter $\hat{\sigma}$. It is typically only weakly biased toward larger values (by 3\%) and rises significantly 
when the photometric noise becomes comparable to the variability amplitude (for faint objects).

\begin{figure*}
\centering
\includegraphics[width=6.8cm]{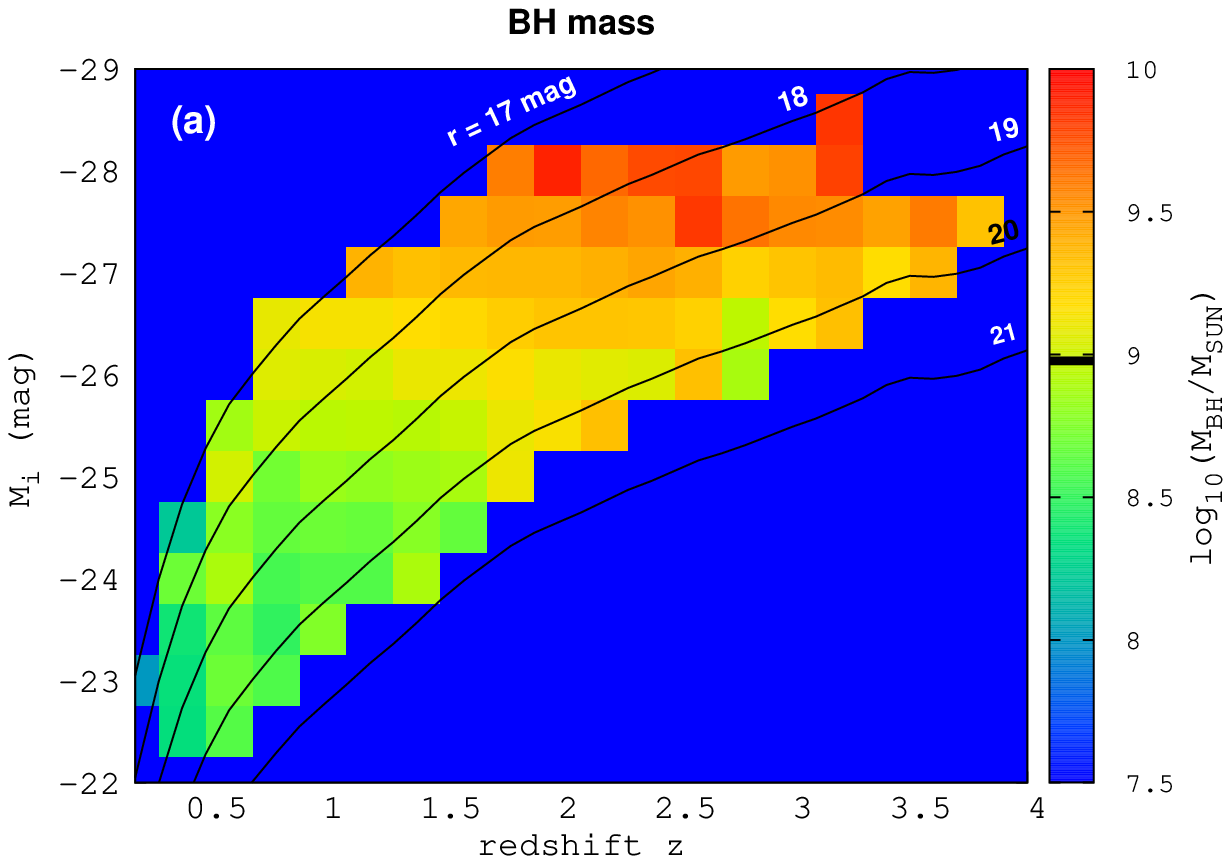} \hspace{0.1cm}
\includegraphics[width=6.8cm]{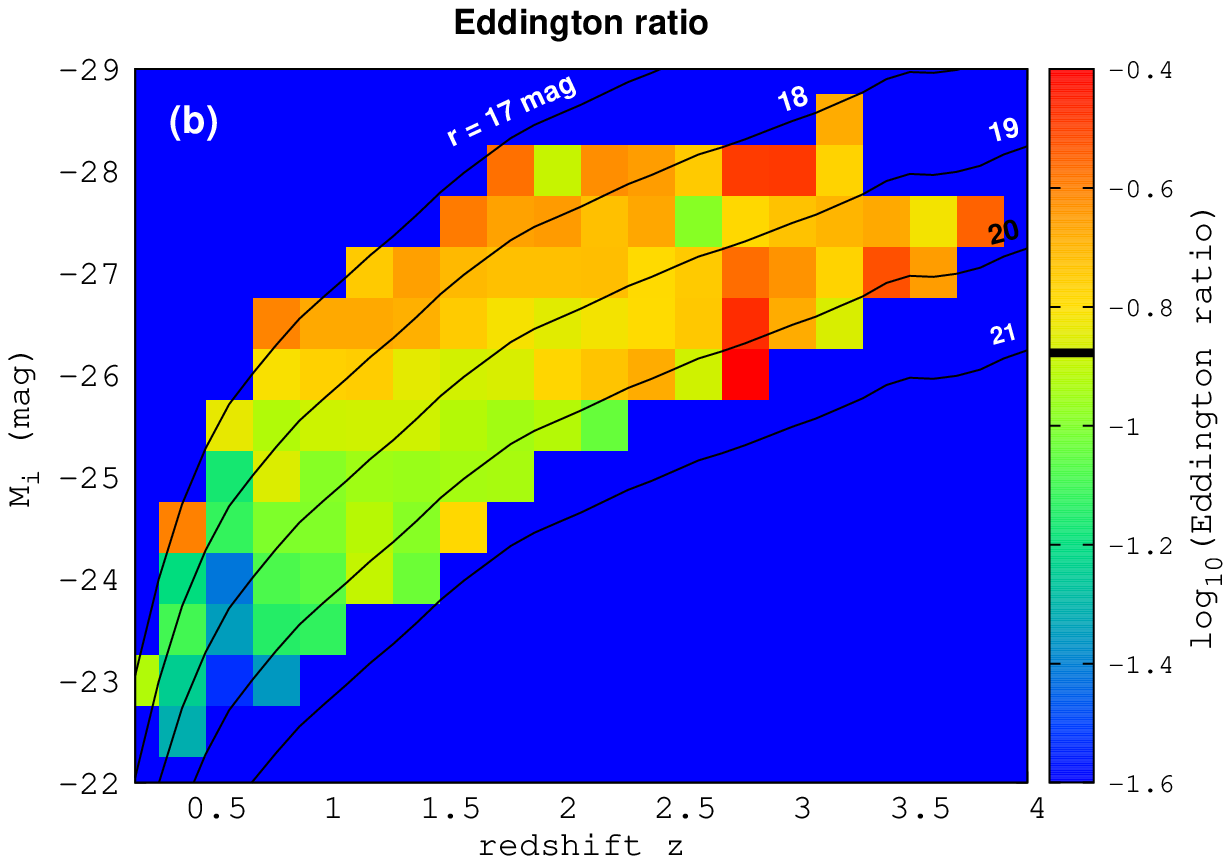}\\
\vspace{0.3cm}
\includegraphics[width=6.8cm]{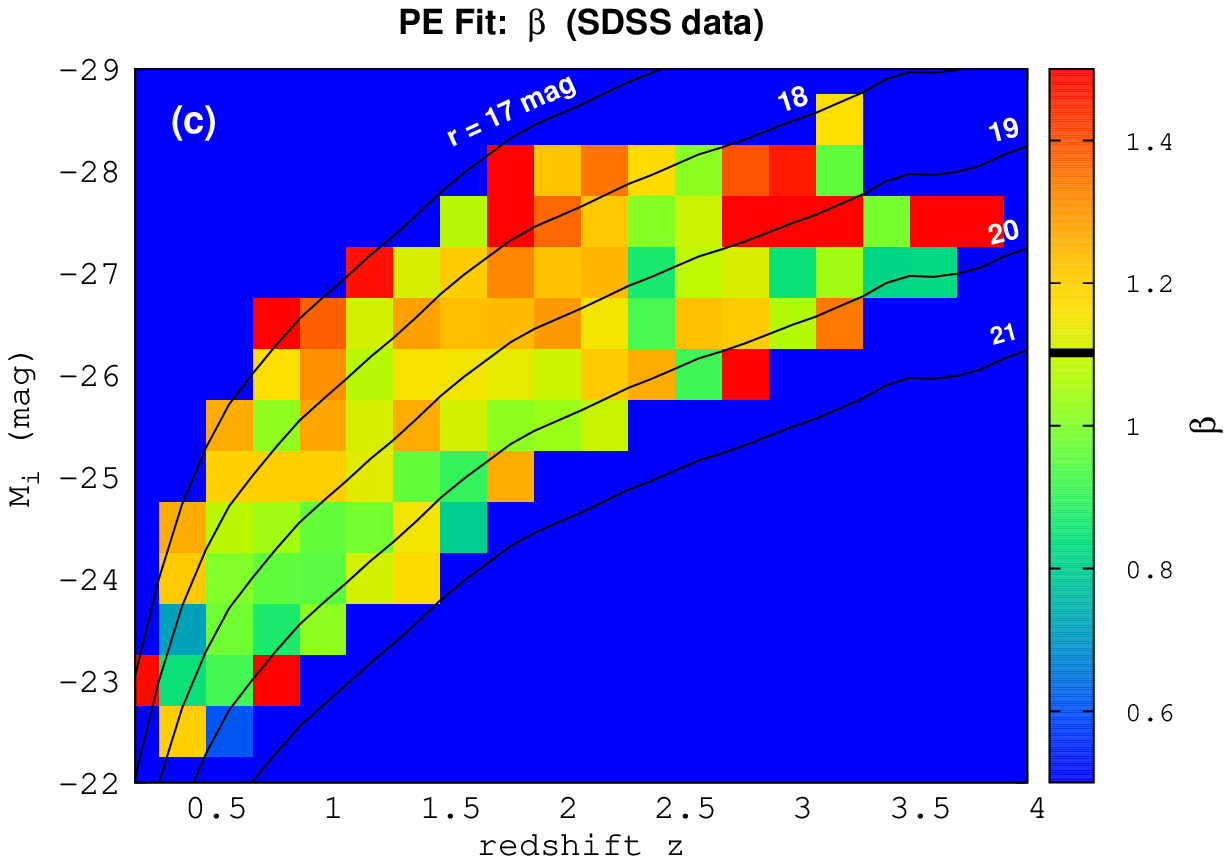} \hspace{0.1cm}
\includegraphics[width=6.8cm]{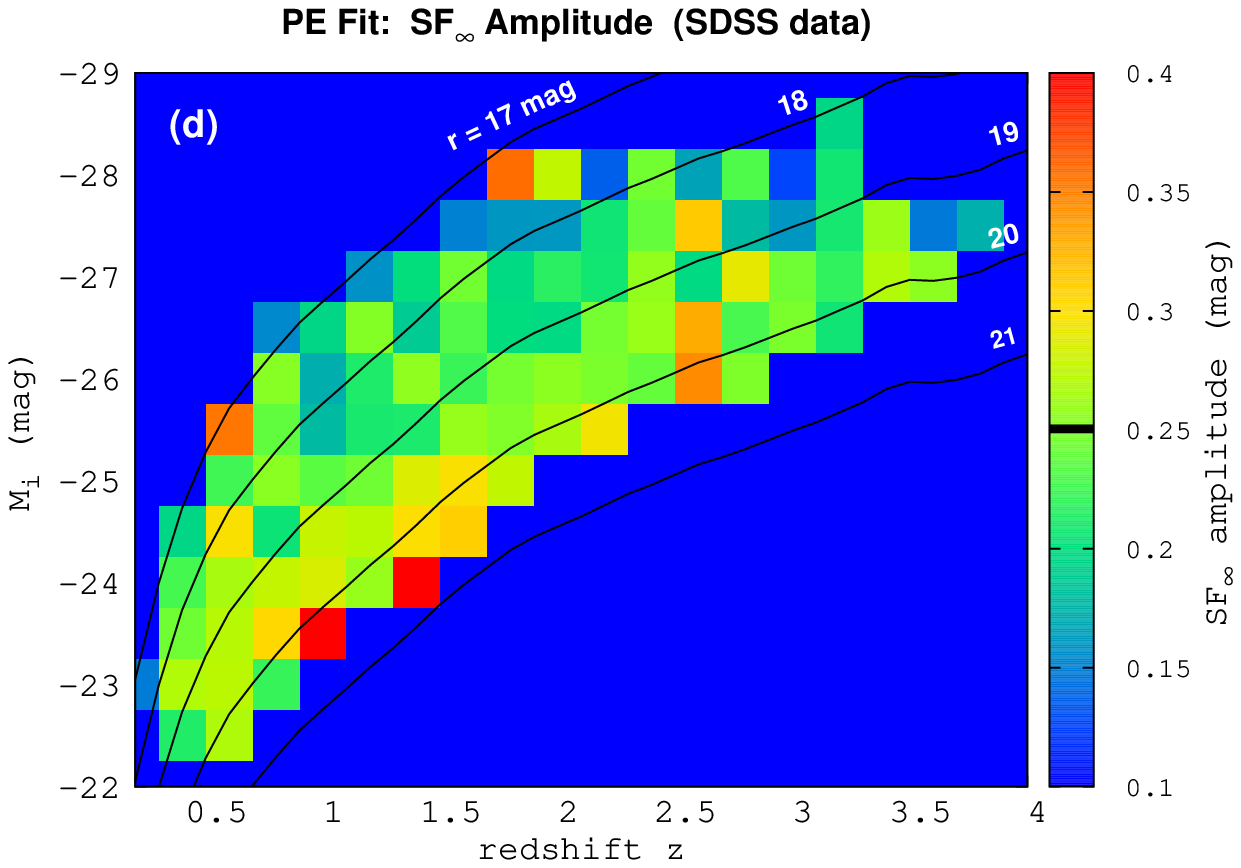}\\
\vspace{0.3cm}
\includegraphics[width=6.8cm]{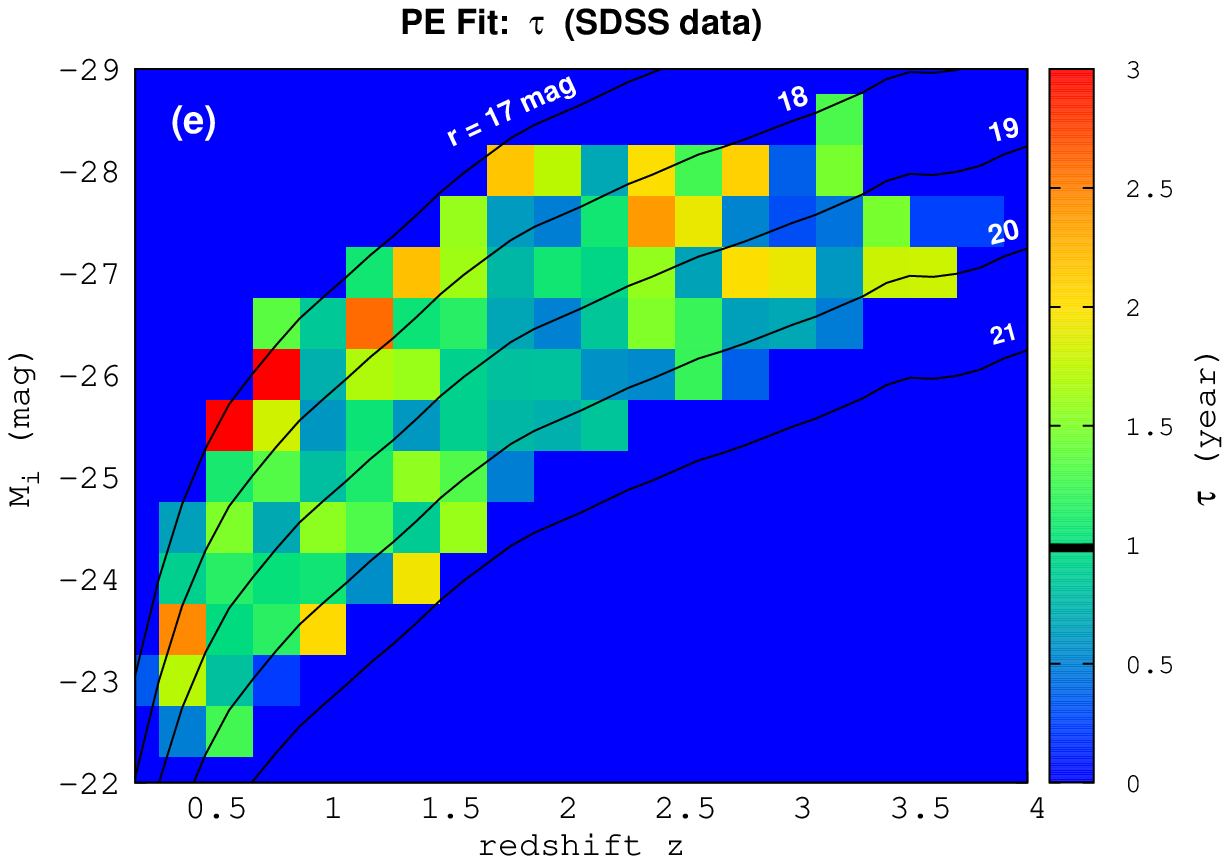} \hspace{0.1cm}
\includegraphics[width=6.8cm]{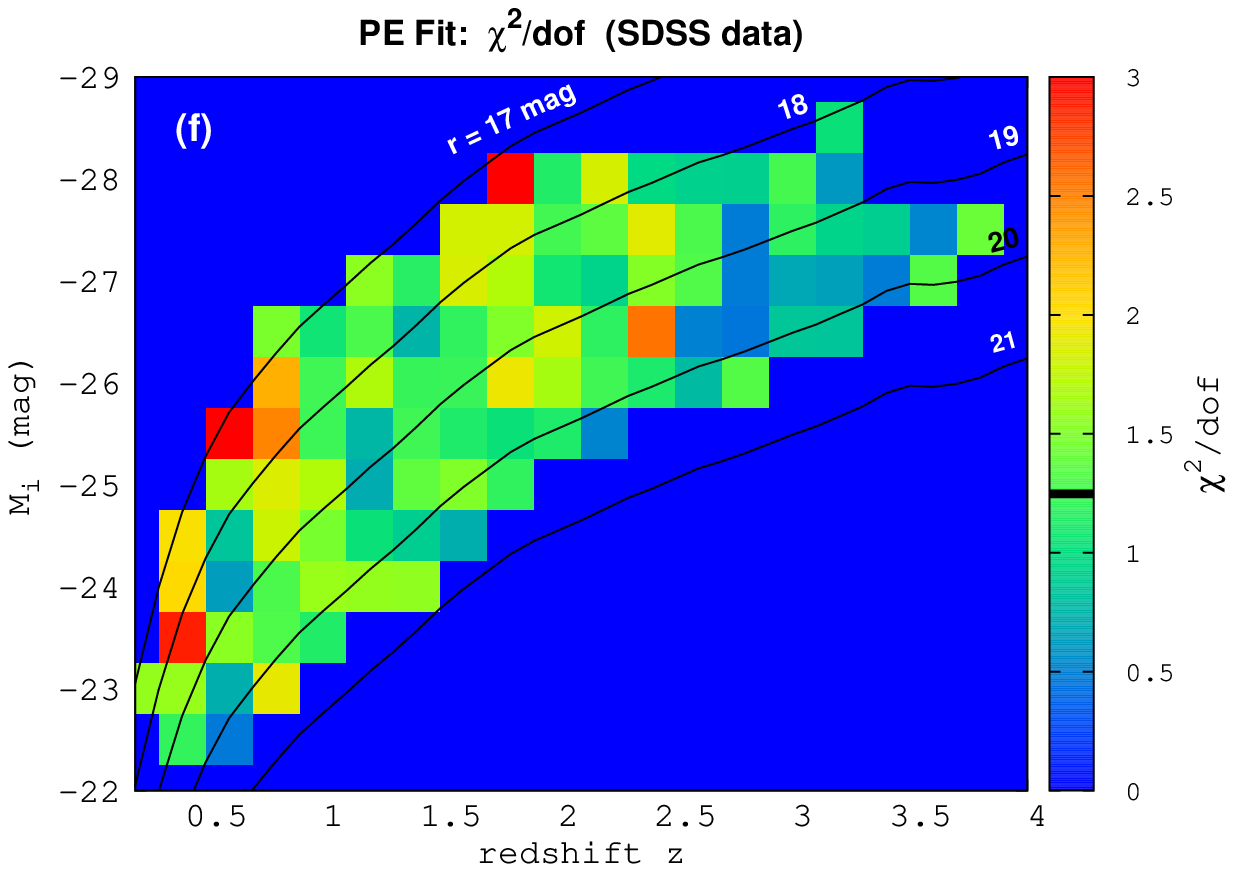}\\
\vspace{0.3cm}
\includegraphics[width=6.8cm]{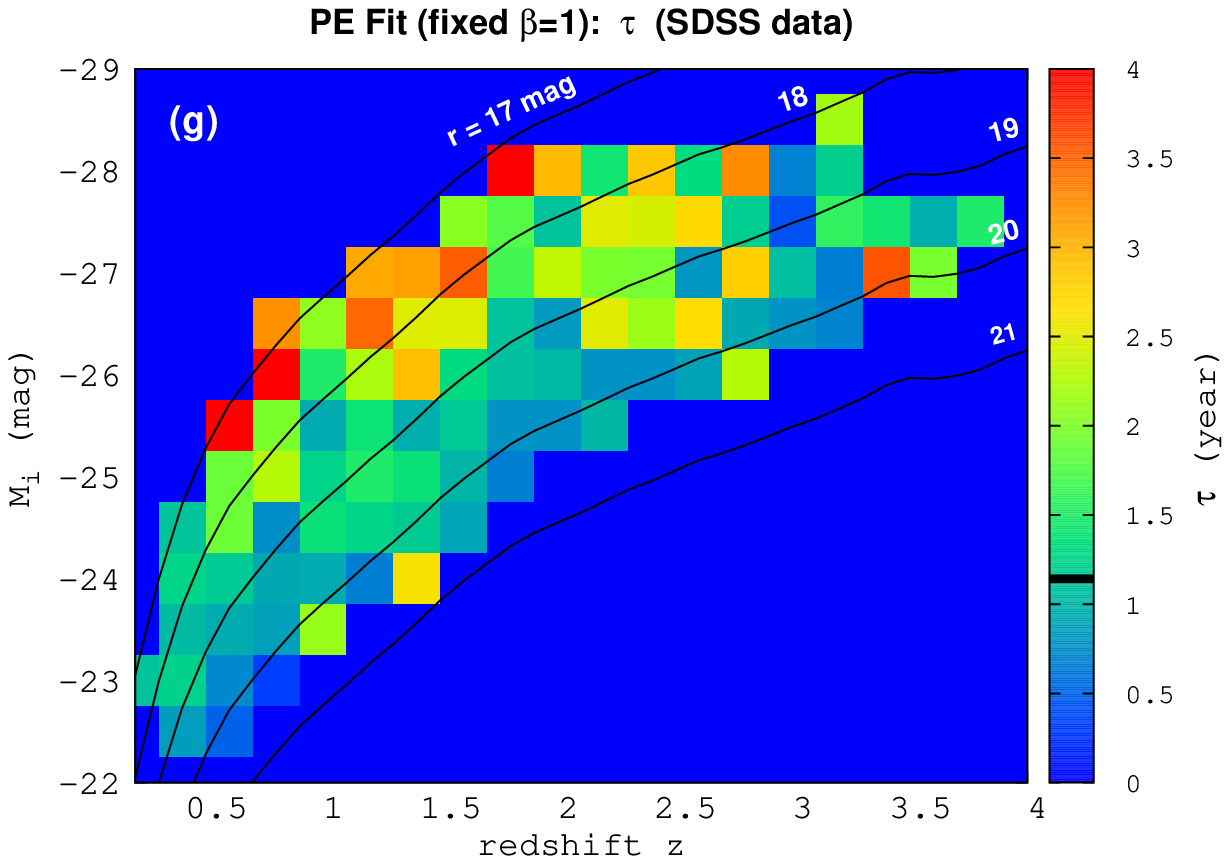} \hspace{0.1cm}
\includegraphics[width=6.8cm]{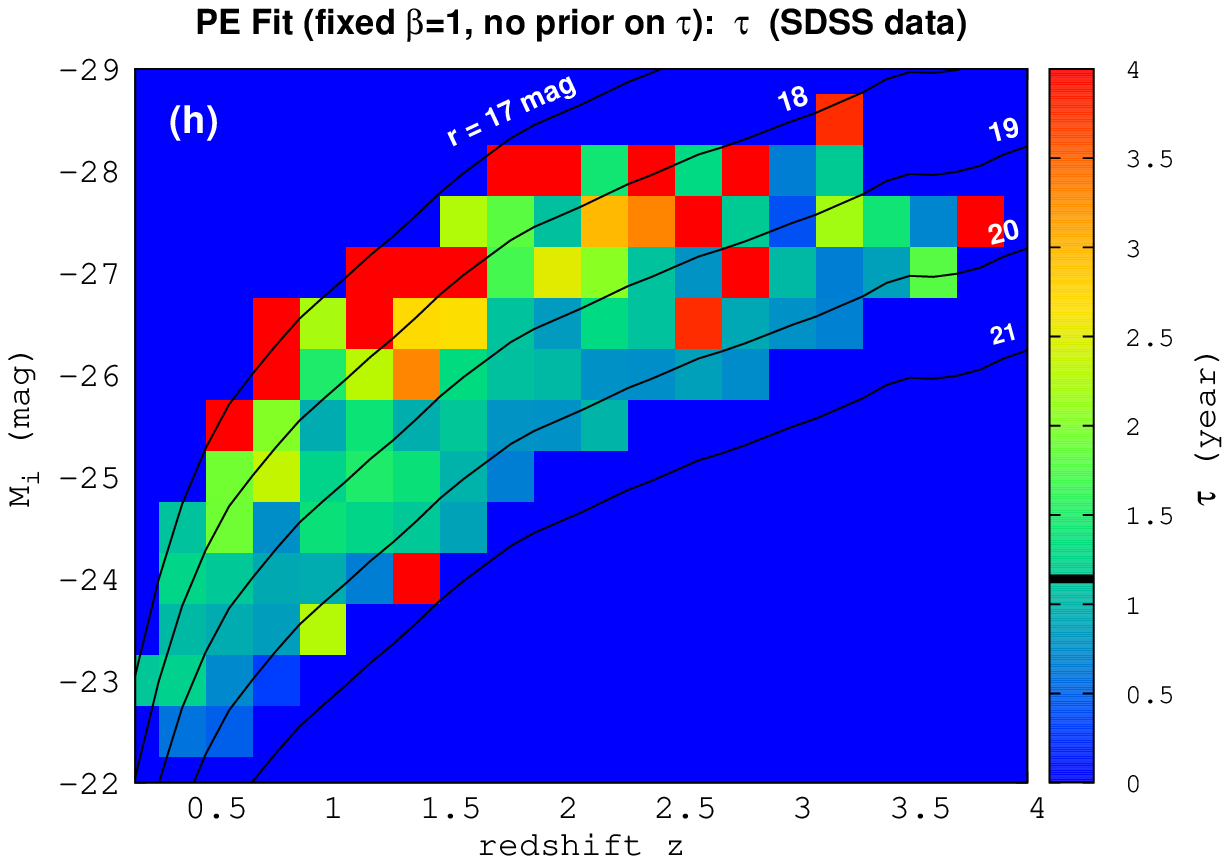}
\caption{Variability analysis of $\sim$9000 SDSS Stripe82 AGNs in the $r$-band. All panels show the same redshift--absolute magnitude ranges and 
also the lines (black) of the constant observed $r$-band magnitude. In panels (a) and (b), we present the mean black hole mass 
and the mean Eddington ratio, respectively. We fit the full, four-parameter SF, where we model the ACF as the power exponential.
The model is given by Equation~(\ref{eq:fullfit}) and has four parameters: the shape of the ACF $\beta$ (panel (c)), 
the variability amplitude at long time lags $SF_{\infty}$ (panel (d)), the timescale of decorrelation $\tau$ (panel (e)), and the noise term (not shown). 
The goodness of fit is presented in panel (f). We also present the timescale $\tau$ for the three-parameter fit (where the fourth parameter is fixed $\beta=1$) in panels (g) and (h).
Nominally, we use a weak prior on the timescale in the three- and four-parameter fits to avoid infinities; in panel (g) we present the timescale with the prior and 
in panel (h) without the prior. The $\chi^2$/dof for these fits slightly rises with increasing $\tau$ from $\sim$1.5 to 2.0--2.5 (not shown).
Each panel is complemented by a dedicated color scale on its right that spans an appropriate range of the parameter space.
The median values (weighted with the AGN number) are marked on the color bars with the thick black line.}
\label{fig:realPE}
\end{figure*}

\begin{figure*}
\centering
\includegraphics[width=6.8cm]{fig9-10a.eps} \hspace{0.1cm}
\includegraphics[width=6.8cm]{fig9-10b.eps}\\
\vspace{0.3cm}
\includegraphics[width=6.8cm]{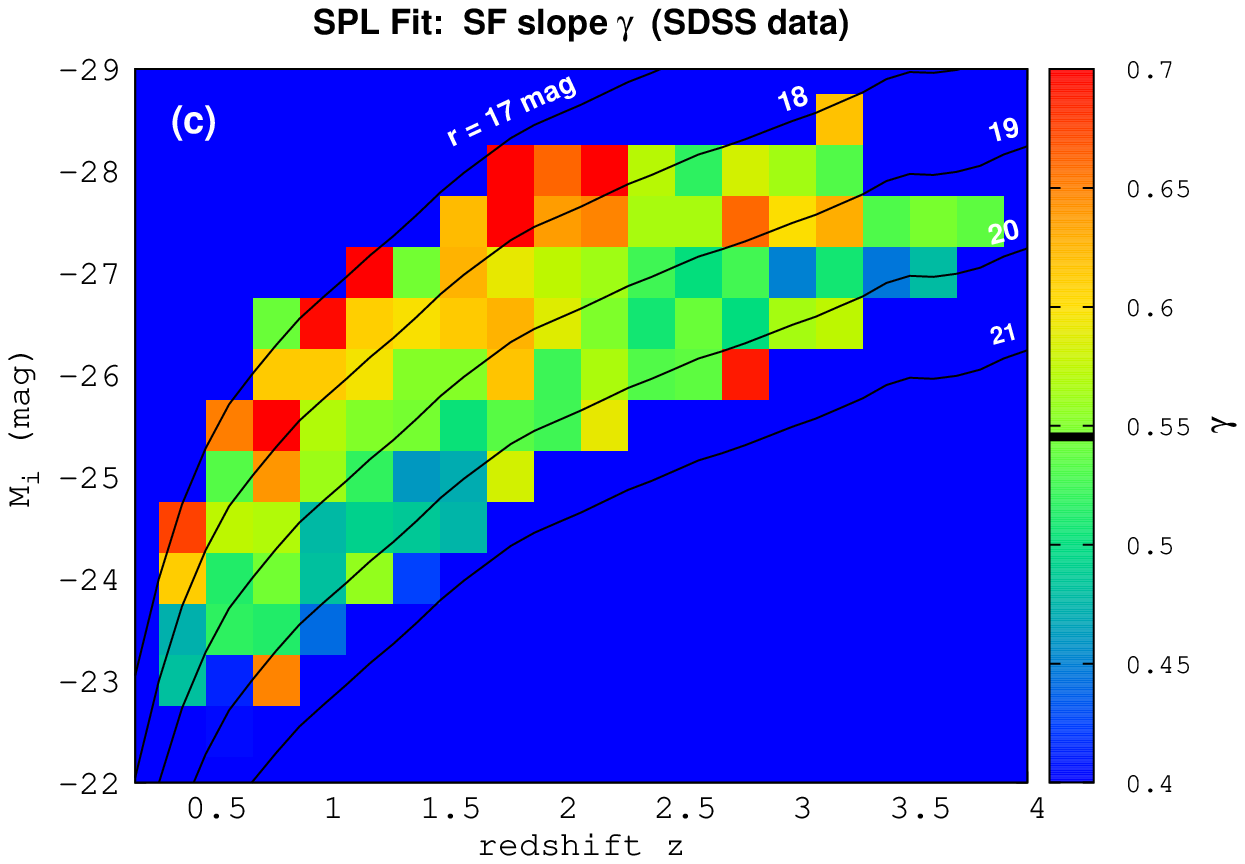} \hspace{0.1cm}
\includegraphics[width=6.8cm]{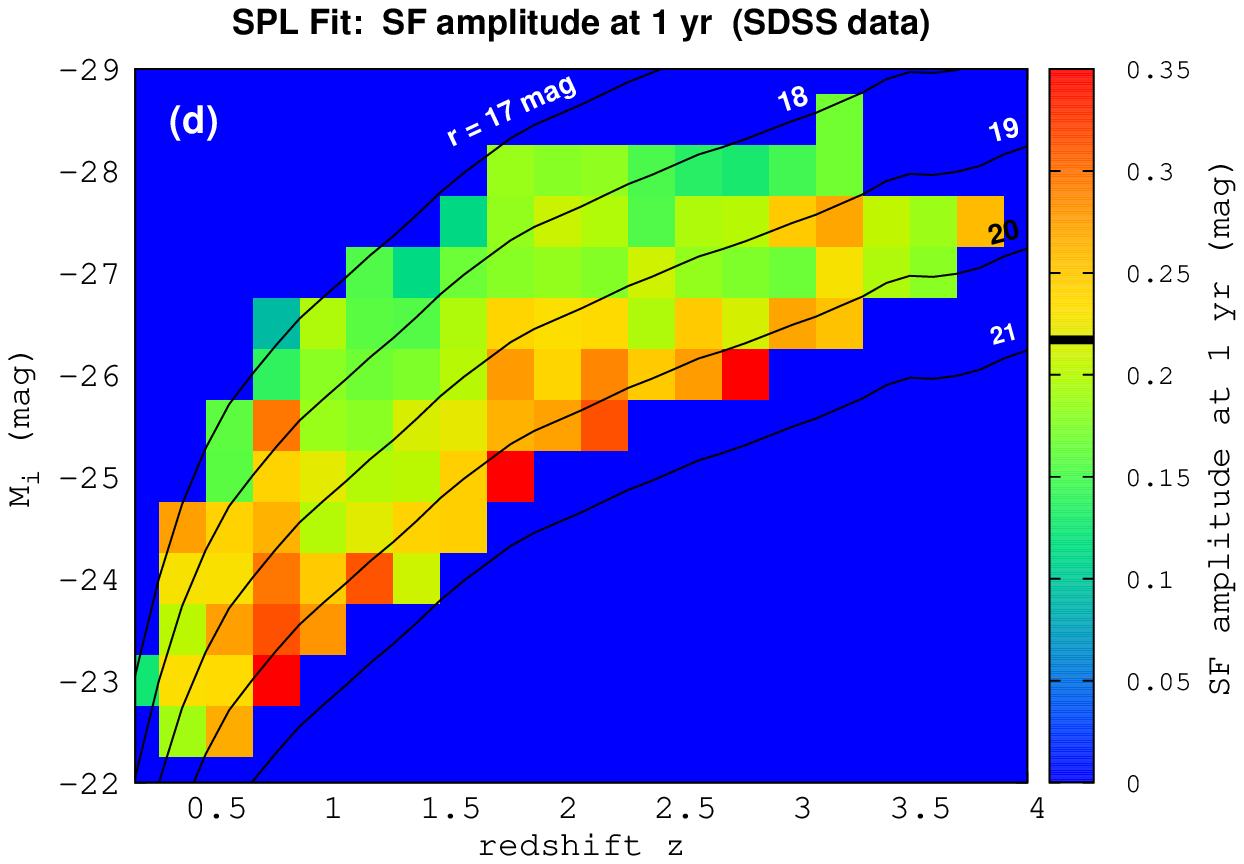}\\
\vspace{0.3cm}
\includegraphics[width=6.8cm]{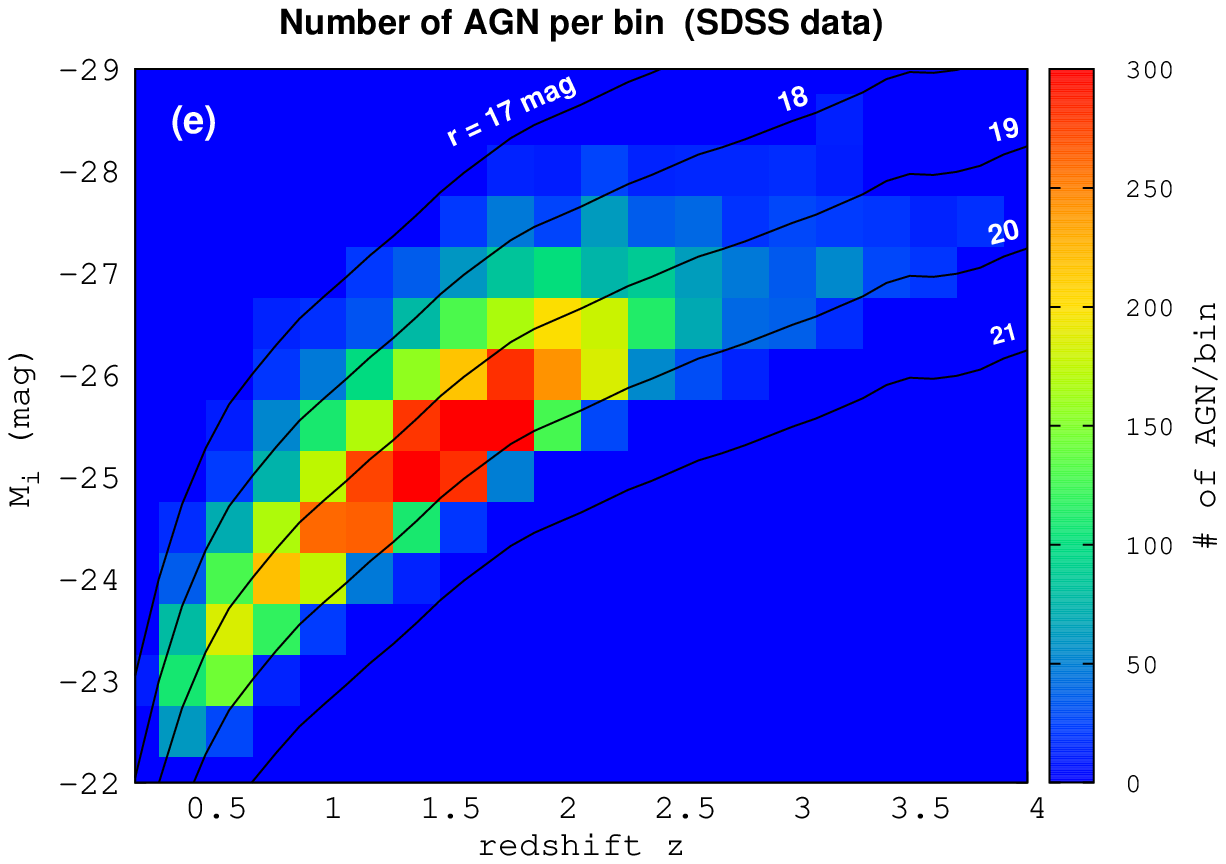} \hspace{0.1cm}
\includegraphics[width=6.8cm]{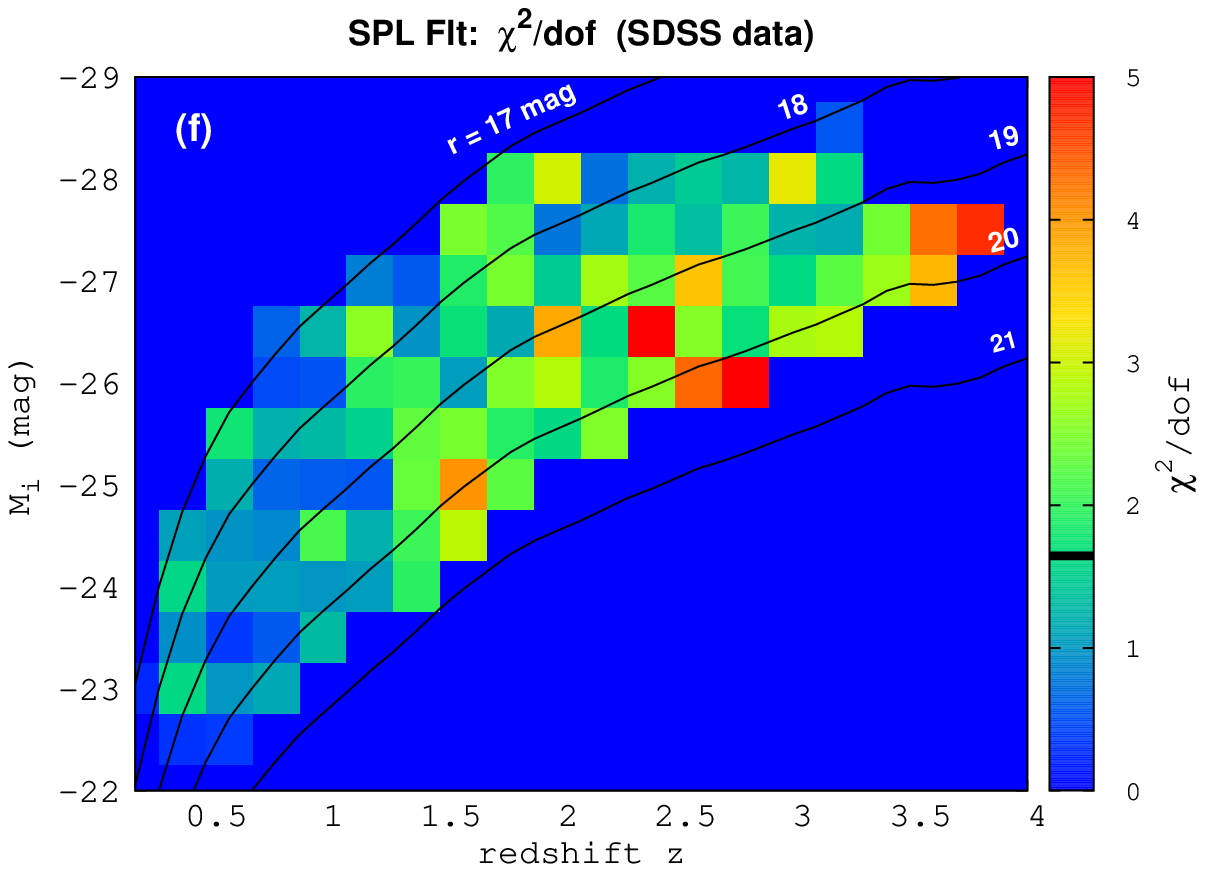}\\
\vspace{0.3cm}
\includegraphics[width=6.8cm]{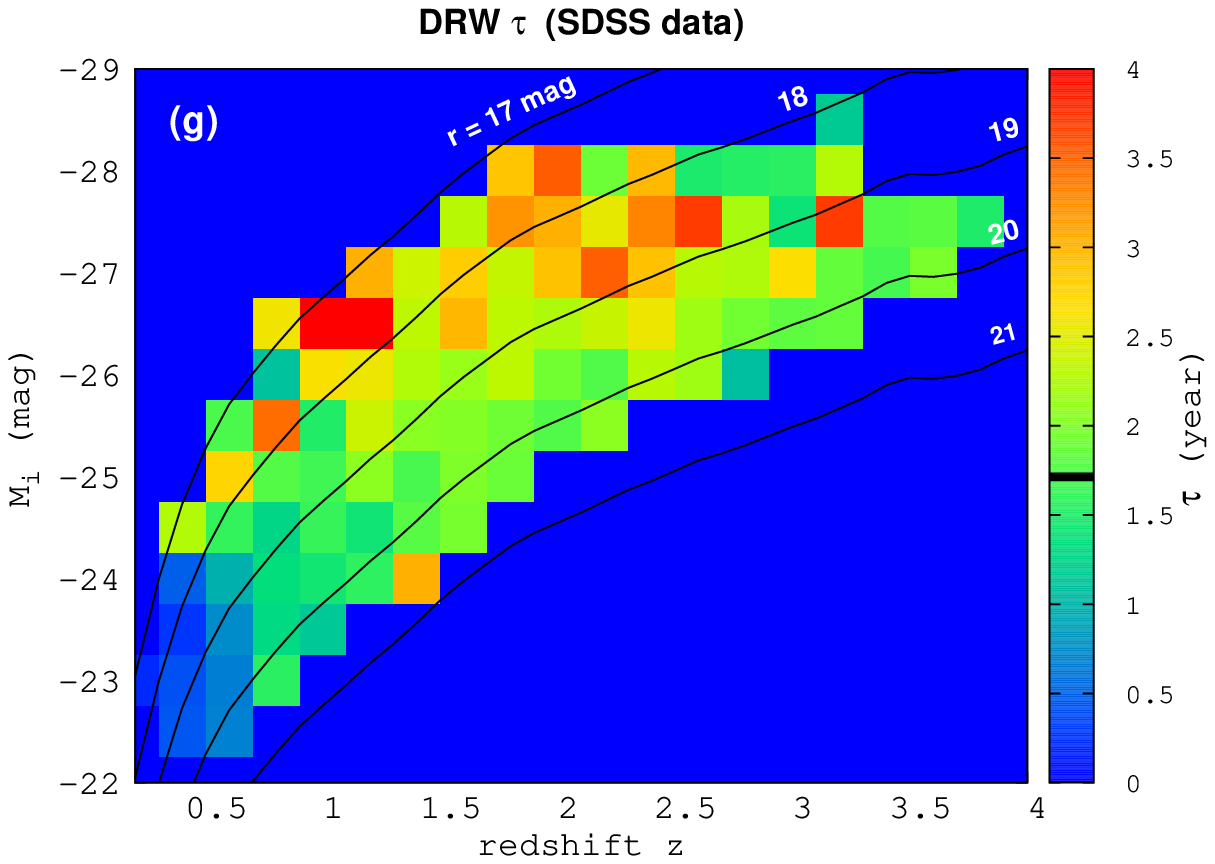} \hspace{0.1cm}
\includegraphics[width=6.8cm]{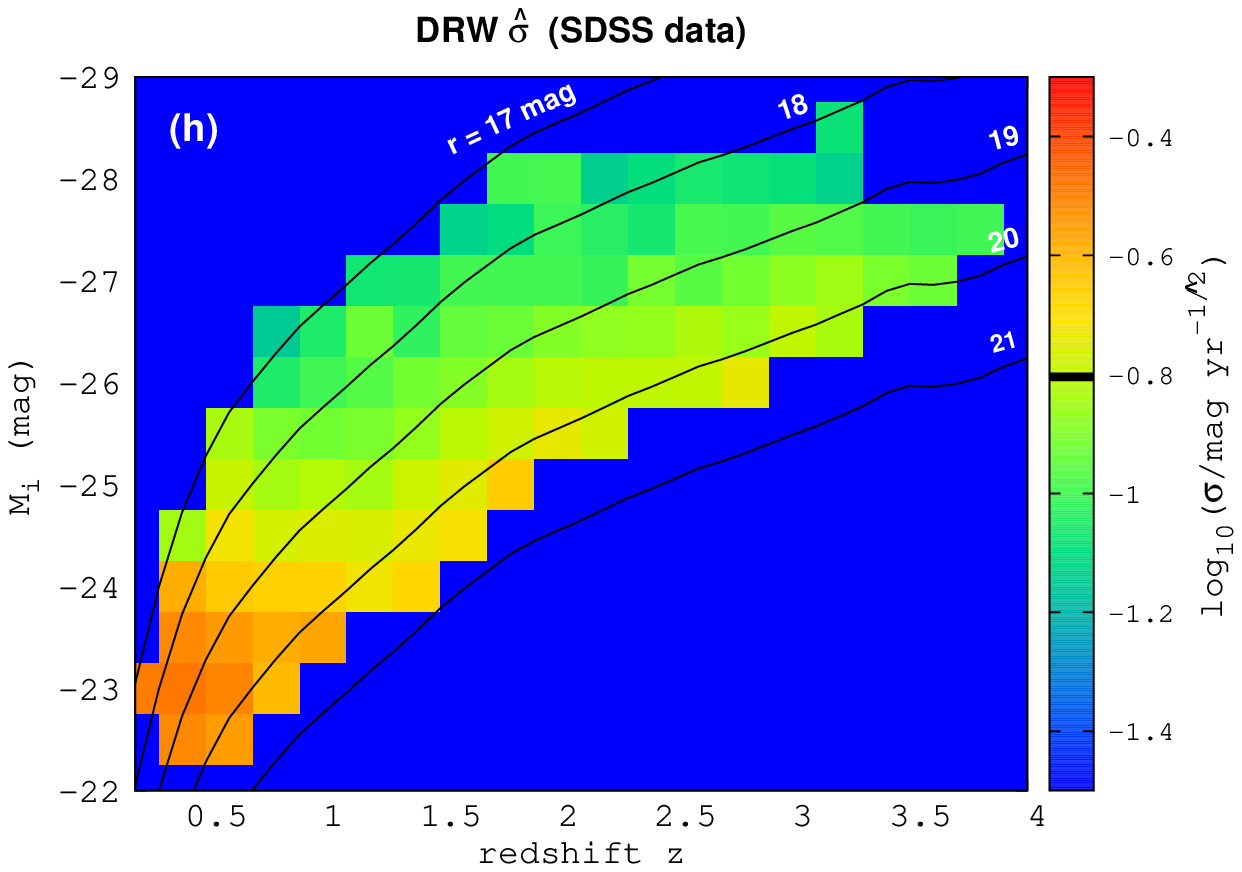}
\caption{Variability analysis of $\sim$9000 SDSS Stripe82 AGNs in the $r$-band. All panels show the same redshift--absolute magnitude ranges and also the lines (black) 
of the constant observed $r$-band magnitude. In panels (a) and (b), we present the mean black hole mass 
and the mean Eddington ratio, respectively. We fit the two-parameter SPL SF (Equation~(\ref{eq:sfrmsMcL2})) in a range $4<\Delta t<365$ days and in panels (c) and (d) show
the resulting parameters $\gamma$ and the amplitude at 1 year, respectively. 
In panel (e), we show the number of AGNs per bin, while in panel (f) we present the $\chi^2$/dof for the SF fit.
We also model each light curve with DRW and present the mean
timescale $\tau$ and the mean modified amplitude $\hat{\sigma}$, shown in panels (g) and (h), respectively. 
They are correlated with the SF $\gamma$ and the amplitude at one year, respectively.
Each panel is complemented by a dedicated color scale on its right that spans an appropriate range of the parameter space.
The median values (weighted with the AGN number) are marked on the color bars with the thick black line.}
\label{fig:realSPL}
\end{figure*}

\section{Discussion}
\label{sec:discussion}

We identified a number of SF definitions in the literature that seem to be flawed, so the reported correlations of variability with the physical parameters that are based on them may also need a revision. In particular, is the amplitude really correlated with the luminosity, rest-frame wavelength, or redshift, or is it simply a correlation with the (partially subtracted) photometric noise? We inspect some of the basic correlations below.

Because in simulations we observe no correlations of the SF parameters with redshift
or the absolute magnitude (Figure~\ref{fig:simulPE} and \ref{fig:simulSPL}),
the real data (Figure~\ref{fig:realPE} and \ref{fig:realSPL}) should provide unbiased measures of these parameters.

\subsection{Correlations}

To untangle dependencies of the variability parameters on the physical parameters, we will consider two cases:
(1) a subsample of AGNs with the fixed luminosity and (2) a subsample of AGNs with the fixed black hole mass.

(1) We selected a sample of 518 AGNs with $19.5<r<20.0$ mag (narrow luminosity range) 
and $1.3<z<1.7$ (similar redshift/emission wavelength) and divided them into five bins in $\eta_{\rm Edd}$, each containing over a hundred AGNs.
We fitted the subensemble SFs with the full SF fit (method (1)) and found a lack of correlation of $\beta$ with the Eddington ratio 
or the black hole mass (Table~\ref{tab:correl}), 
but we found an anticorrelation of the asymptotic amplitude with the Eddington ratio of the form $\log(SF_\infty)\propto(-0.28\pm0.06)\log(\eta_{\rm Edd})$, in agreement with the value found by \cite{2010ApJ...721.1014M}, that is, $-0.23\pm0.03$, and in rough agreement with \cite{2008MNRAS.383.1232W}, who measured $\sim-0.15$.  Please note, however, that they used the incomplete SF equation from \cite{1996ApJ...463..466D}.

Because we are inspecting here a narrow luminosity range and $-1.5<\log(\eta_{\rm Edd})<0$, the change in
the Eddington ratio ($\eta_{\rm Edd}\propto L_{\rm bol} M_{\rm BH}^{-1}$) is due to the changing black hole mass in a range $8<\log(M_{\rm BH}/M_\odot)<9.5$. 
We observe a correlation of the amplitude with the black hole mass of the form $\log(SF_\infty)\propto(0.29\pm0.04)\log(M_{\rm BH})$,
slightly higher than that (0.18) found in \cite{2010ApJ...721.1014M}.

The timescale $\tau$ is correlated with the black hole mass $\log(\tau)\propto(0.38\pm0.15)\log(M_{\rm BH})$,
with a somewhat higher index (0.21) than in \cite{2010ApJ...721.1014M},
and is anticorrelated with the Eddington ratio $\log(SF_\infty)\propto(-0.36\pm0.15)\log(\eta_{\rm Edd})$ (Table~\ref{tab:correl}).

(2) We selected 837 AGNs in a range $8.7<\log(M_{\rm BH}/M_\odot)<9.3$ (constant black hole mass) and $1.3<z<1.7$ (similar redshift/emission wavelength)
and divided them into six luminosity classes with each containing 50--200 AGNs. We fitted the sub-ensemble SFs with the full SF fit (method (1)).
We observe a slight increase of $\beta$ with the increasing bolometric luminosity (an SPL index 0.10) 
and the Eddington ratio (0.16). The asymptotic variability $\log(SF_\infty)\propto(-0.35\pm0.05)\log(L_{\rm bol})$ and 
$\log(SF_\infty)\propto(-0.55\pm0.06)\log(\eta_{\rm Edd})$  (Table~\ref{tab:correl}).
The timescale is nearly independent of the bolometric luminosity and the Eddington ratio,
$\log(\tau)\propto(-0.05\pm 0.17)\log(L_{\rm bol})$ and $\tau\propto(-0.16\pm0.25)\log(\eta_{\rm Edd})$, respectively,
and is consistent with being constant as reported by \cite{2010ApJ...721.1014M}, while staying in contrast to the positive correlation in \cite{1993Natur.366..242H}.

\begin{figure*}
\centering
\includegraphics[width=6.8cm]{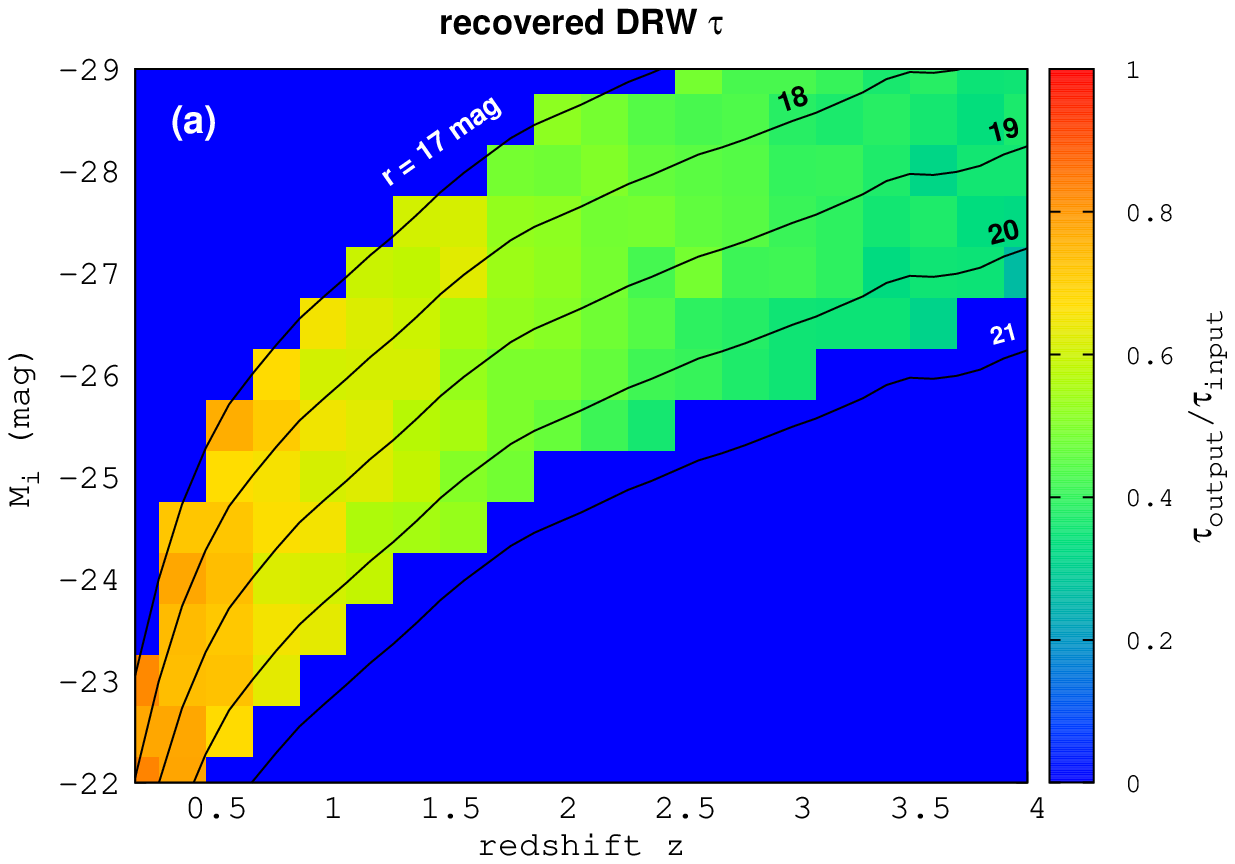} \hspace{0.1cm}
\includegraphics[width=6.8cm]{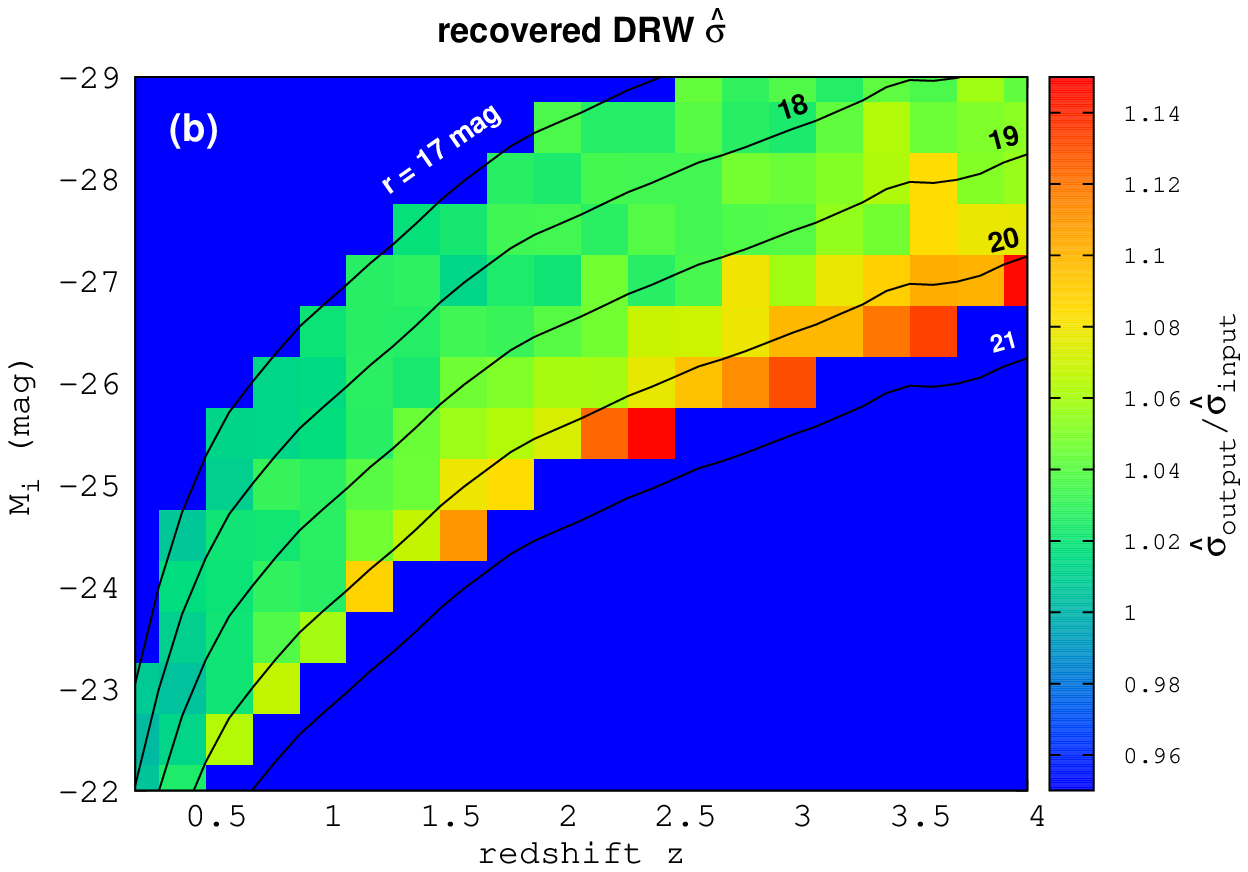}\\
\vspace{0.3cm}
\caption{DRW simulations (with $\tau=500$ days and $SF_\infty=0.18$ mag) and modeling of a 1000 light curves per bin. 
Ratios of the output to input parameters are shown. The returned timescales (panel (a)) are increasingly underestimated 
with increasing redshift (in fact with the increasing ratio of the timescale to the experiment length),
and there is only a weak overestimate ($\sim$3\%) of $\hat{\sigma}$ for $r<19.5$ mag (panel (b)), slightly increasing with redshift.
The longer the experiment span, the more reliable the estimate of the true DRW timescales. However, any recovered DRW parameters should be debiased by simulation means.}
\label{fig:simulDRW}
\end{figure*}

\begin{table}
\begin{center}
\caption{Correlations of the AGN Variability and Physical Parameters\label{tab:correl}}
\begin{tabular}{lcr}
\tableline\tableline
Variability Parameter & Physical Parameter & Power-law Index\\
\tableline \\
\multicolumn{3}{c}{Constant Luminosity} \\
\tableline
$\log(\beta)$     & $\log(M_{\rm BH})$         & $ 0.01 \pm 0.05$ \\
$\log(\beta)$     & $\log(\eta_{\rm Edd})$     & $-0.01 \pm 0.04$ \\
$\log(SF_\infty)$ & $\log(M_{\rm BH})$         & $ 0.29 \pm 0.04$ \\
$\log(SF_\infty)$ & $\log(\eta_{\rm Edd})$     & $-0.28 \pm 0.06$ \\
$\log(\tau)$      & $\log(M_{\rm BH})$         & $ 0.38 \pm 0.15$ \\
$\log(\tau)$      & $\log(\eta_{\rm Edd})$     & $-0.36\pm 0.15$ \\
\tableline \\
\multicolumn{3}{c}{Constant Black Hole Mass} \\
\tableline
$\log(\beta)$     & $\log(L_{\rm bol})$    & $ 0.10 \pm 0.03$ \\
$\log(\beta)$     & $\log(\eta_{\rm Edd})$ & $ 0.16 \pm 0.05$ \\
$\log(SF_\infty)$ & $\log(L_{\rm bol})$    & $-0.35 \pm 0.05$ \\
$\log(SF_\infty)$ & $\log(\eta_{\rm Edd})$ & $-0.55 \pm 0.11$ \\
$\log(\tau)$      & $\log(L_{\rm bol})$    & $-0.05 \pm 0.17$ \\
$\log(\tau)$      & $\log(\eta_{\rm Edd})$ & $-0.16 \pm 0.25$ \\
\tableline
\end{tabular}
\tablecomments{For $\Delta t\ll \tau$ the slope of the SF $\gamma$ is related to the power $\beta$ as $\beta\equiv 2\gamma$, 
hence relations for $\beta$ can be directly translated to SF slopes.}
\end{center}
\end{table}

\subsection{Comments on DRW}

A number of studies (e.g., \citealt{2012ApJ...753..106M,2013ApJ...765..106Z,2016ApJ...817..119K}) based on {\it Kepler} results from \cite{2011ApJ...743L..12M} 
stated that DRW is not the right model for ``very short timescales'' that are not or are weakly probed by ground-based surveys.
This is only partly true; namely, {\it Kepler} PSDs appear to be steeper, but the frequencies involved, $10^{-8}$--$10^{-5}$ Hz or 1.2--115 days, are in 
fact well probed by ground-based studies. In particular, the OGLE survey with the cadence 2--4 days (\citealt{2015AcA....65....1U})
has been the gold standard for many astrophysical fields for over two decades, and \cite{2013ApJ...765..106Z} was unable to find deviations from DRW in such data.
Note, however, that \cite{2016MNRAS.459.2787K} recently showed that direct modeling of light curves with DRW is equally good for both DRW and non-DRW processes and cannot be used to identify what process is being modeled. 
There is rather a full discrepancy between {\it Kepler} and the ground-based surveys, than the claimed 
one on ``very short time-scales'' (\it Kepler} PSDs hit the white noise 
at (or below) the 1--day cadence, so it does not probe the covariance of the signal below these timescales). 
So \cite{2015MNRAS.453.2075K} justly asked a question if {\it Kepler}'s light curves need reprocessing, 
and they positively answered to this question. In fact, steeper PSDs or SFs mean a faster variability that is ``easy to achieve'' 
by improper photometric procedures, such as when too much light/background is subtracted from a light curve.

\cite{2011ApJ...730...52K} considered AGN variability in light of a linear combination of multiple OU processes. Such a combination
of DRW processes may lead to an average ACF with powers different from $\beta=1$, although individual processes are DRW ($\beta=1$). 
So, for example, a shallower or steeper SF may mean either one process that is different from DRW or a combination of DRW processes leading to 
shallower or steeper SFs.

We will now discuss a DRW modeling of the simulated SDSS light curves with constant $\tau=500$ days and $SF_\infty=0.18$ mag, but for a range of redshifts and absolute magnitudes.
For each redshift--absolute magnitude bin there are 1000 AGN light curves generated and modeled with DRW (Figure~\ref{fig:simulDRW}).
Each model parameter is represented by a posterior probability distribution, where the peak is identified here with the best value and the width of the distribution (the parameter uncertainty)
is not considered. Of the 1000 best parameter values per bin we calculate their median.
From Figure~\ref{fig:simulDRW}, we see that the measured DRW parameters are a strong function of the ratio of the input timescale (which increases with redshift)
to the total experiment span. What happens
if you cut an AGN light curve, let us say, in half? Well, because the ratio of $\tau$ to the experiment length increases, the returned timescale decreases.
So a longer light curve, with the same underlying process as the shorter one, will have a longer (and closer to the true value) timescale than the cut light curve. Because this was not accounted for,
it could have been the reason for problems in finding time lags due to light time travel in ``the photometric reverberation mapping'' 
of the OGLE-III (eight years long) and OGLE-IV (then four years long) light curves in \cite{2016ApJ...819..122Z}.

\subsection{Typical AGN variability}

In Figures~\ref{fig:realPE} and \ref{fig:realSPL}, the SDSS S82 data were divided into bins primarily to avoid mixing AGNs with
the different photometric noise, which causes problems in subtracting or modeling it, but also to investigate 
any possible correlations of variability with the physical parameters of AGNs. 
We will now combine the results from these figures to obtain ``typical'' AGN variability parameters.

Because bins contain different numbers of AGNs (panel (e) in Figure~\ref{fig:realSPL}), imagine a simple situation where 
we consider only two bins, one with 10 AGNs with an ``incorrect'' SF value
and the second bin with 300 AGNs and the ``correct'' SF value. If they had equal weights, then the average of the two would be off from the ``correct value.''
Therefore we will add weights in averaging bins: when constructing histograms, we simply count each bin a number of times equal to the number of AGNs in it.

\begin{figure}
\centering
\includegraphics[width=8.2cm]{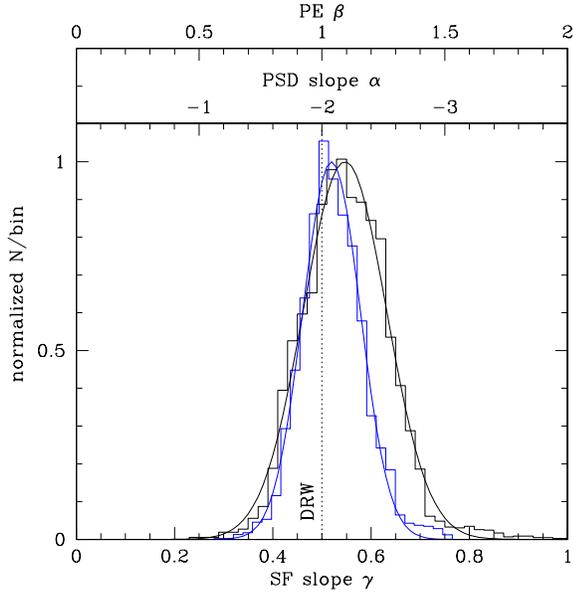}
\caption{Histograms, weighted with the number of AGNs in a bin and normalized so the Gaussian fits peak at 1, of SF slopes $\gamma$ (PE $\beta$ and PSD slope $\alpha$ on top) 
from fitting the full SF function (black) and an SPL SF for $4<\Delta t<365$ days (blue) 
to the redshift--absolute magnitude bins (corrected for biases). The histograms can be modeled relatively well as Gaussians with the mean values $\gamma=0.55$ and 0.52, 
and dispersions $\sigma_\gamma=0.08$ and 0.06, respectively. The DRW model ($\gamma=0.5$) is marked with the vertical dotted line.}
\label{fig:histGamma}
\end{figure}

In Figure~\ref{fig:histGamma}, we present weighted histograms of the SF slope $\gamma$. They were obtained from the two fitting methods, 
the full-fit SF slopes returning $\beta\equiv2\gamma$, and the SPL SF fits. 
These histograms peak at $\gamma=0.55\pm 0.08$ and $0.52\pm 0.06$, respectively. These values are a little higher than the expected $\gamma=0.5$ from DRW but are entirely consistent with it.
The timescale $\tau$ has a wide and non-Gaussian distribution. The median value is $\tau=0.98$ year, and we estimate the error bar using the IQR; it is $\sigma_\tau=0.41$ year.
If we fitted this histogram with a Gaussian, it yields $\tau=0.97\pm0.46$ year. 
The weighted histogram for the asymptotic amplitude is $SF_\infty=0.25\pm0.06$ mag, while the one for the amplitude at one year is $SF_0=0.22\pm0.06$ mag. Note, however, that the SFs are already slightly flattening at this time lag, so the latter value is simply overestimated. The true value is $SF_0=0.20$ mag.

\section{Summary}
\label{sec:summary}

In this paper, we have reviewed basic variability observables that are often used in optical and infrared AGN studies. 
From basic principles we show how to correctly measure the SF so it can be directly linked to the 
covariance function, or equivalently the autocorrelation function, of the underlying stochastic process (or processes) causing the variability
(please refer to \citealt{2009ApJ...698..895K,2011ApJ...730...52K,2014ApJ...788...33K} for details on the stochastic processes).

Prior to the measurement of the AGN variability parameters from the SDSS S82 data, we simulate similar data sets where the input parameters are known.
By comparison of the output and input parameters, we investigate biases and systematics that can potentially be present in the SF measurement from the real data.
Also by simulation means we review commonly used SF measures and point out their problems.
The most common one is the subtraction of an incomplete or no noise term, leading to flat SF slopes.
SFs can be successfully modeled as an SPL for the rest-frame time lags in the range between days and about one year (before turning into white noise),
provided that the noise term is subtracted off carefully (and fully).

The key result of this paper is the slope of the SF measured from two methods:
the full SF modeling using the PE covariance matrix and the SPL SF fitting in the ``red noise regime.'' 
The two methods produce SF slope distributions peaking at $\gamma=0.55\pm 0.08$ and $0.52\pm 0.06$, respectively, which are
slightly steeper than but consistent with a single DRW model (Figure~\ref{fig:histGamma}). The 
covariance function (or the auto correlation function) of the underlying stochastic process leading to variability is well described by the exponential
$\exp(-(\Delta t/\tau)^\beta)$, where $\beta=2\gamma$ is in the vicinity of unity. Of the theoretically tested scenarios on the origin of variability with the available predictions
on SFs, the most viable is the model of accretion disk instabilities with $\beta\approx0.9$ (\citealt{1998ApJ...504..671K}).
Because SFs are model-independent and ``raw'' measures of variability, in fact we provide, for the first time,
 the direct proof that AGN variability is akin to the DRW process. The caveat is, however, that other covariance functions not considered here may potentially lead to similar 
SF slopes. \cite{2016MNRAS.459.2787K} simulated both DRW and non-DRW AGN light curves, modeled them as the DRW process, and found that
it is so flexible that returns equally good fits for both DRW and non-DRW stochastic processes.
Hence, light curve modeling with DRW cannot be used as a proof for the variability being caused by the DRW process.
In fact, in \cite{2010ApJ...708..927K} we had already shown that DRW can correctly model 
deterministic processes (i.e., nonstochastic) such as pulsating variable stars, where the timescale is identified with the variability period.
Whether DRW is the only underlying process that we observe is not clear. While the distribution of the SF slopes peaks at $\gamma\approx0.5$, as expected for DRW, we also observe
a weak steepening of the SF for bright AGNs (with the lack of correlation with the black hole mass or Eddington ratio). 
This effect is observed both in the SF modeling with freed slopes, but also when fixing $\gamma=0.5$ as an increase of the decorrelation timescale with the increasing luminosity.

The typical SF amplitude at one year is $SF_0=0.22\pm0.06$ mag, but an SPL fit overestimates this value 
because the true SF is already in the pink noise regime (turning over).
The true value of the SF amplitude at one year is $SF_0=0.20$ mag, while the asymptotic variability is $SF_\infty=0.25\pm0.06$ mag.
The asymptotic variability $SF_\infty$ is correlated with the black hole mass (with an SPL index of $0.29\pm0.04$) 
and strongly anticorrelated with the luminosity (with an SPL index of $-0.35\pm0.05$).

The distribution of decorrelation timescales differs from a Gaussian but can be approximated with $\tau=0.97\pm0.46$ year 
(median $\tau=0.98$ year, $\sigma_\tau=0.41$ year obtained from IQR), despite a weak prior on $\tau$ to be 1.5 year added in the minimization procedure.
The timescale is correlated with the black hole mass (with an SPL index of $0.38\pm0.15$) and
does not depend on luminosity (an SPL index of $-0.05\pm0.17$).

AGN variability can be further studied with longer, already existing or near-future data sets. There exists a data set of about 800 AGNs lying behind the Magellanic Clouds and discovered mostly
by the Magellanic Quasars Survey (\citealt{2013ApJ...775...92K}). They have been observed for nearly two decades by the OGLE survey (\citealt{2015AcA....65....1U}) 
and should provide further clues on the decorrelation timescale, as the rest-frame timescales will be $\Delta t>6$ years. 
The advantage of these light curves is twofold: (1) they have a cadence of a few days and were obtained with a single telescope with nearly identical detector setup (filters, pixel scale)
for different phases of the OGLE survey, which will ease and make robust the data analysis; (2) they are viewed through the Magellanic Clouds, 
so a large number of constant stars are available to correctly estimate the photometric noise. 
The {\it GAIA} satellite (\citealt{2001A&A...369..339P}), collecting data since mid-2014, 
is planned to last five years, and it will scan the sky a median of 70 times. It will provide a usable photometry for broad $g<20$ mag for a billion objects, including
hundreds of thousands of AGNs. While these data will weakly probe the SF at the decorrelation timescale (they simply would have to be longer), they will enable
a study of the shape of the ACF from ensembles based on a really large number of AGNs.
The Large Synoptic Survey Telescope (LSST; \citealt{2008arXiv0805.2366I}) will be a 10-year-long, deep $r<24.5$ AB mag survey scanning three quarters of the sky in six 
optical/infrared filters from the southern hemisphere.
Although the typical cadence per filter will be only a month, it will provide an excellent data sets in six filters to study AGN SF/ACFs and their dependence on the physical parameters of AGNs.


\acknowledgments

We are grateful to Chris Kochanek, Andrzej Udalski, and the anonymous referee for reading the manuscript and providing us with comments that 
improved the flow and clarity of the presented arguments.
SK acknowledges the financial support of the Polish National Science Center through the
OPUS grant number 2014/15/B/ST9/00093 and MAESTRO grant number 2014/14/A/ST9/00121.


\end{document}